# On the Origin & Thermal Stability of Arrokoth's and Pluto's Ices

C.M. Lisse[1], L.A. Young[2], D.P. Cruikshank[3], S.A. Sandford[3], B. Schmitt[4], S.A. Stern[2], H.A. Weaver[1], O. Umurhan[3], Y.J. Pendleton[5], J.T. Keane[6], G.R. Gladstone[7], J.M. Parker[2], R.P. Binzel[8], A.M. Earle[8], M. Horanyi[9], M. El-Maarry[10], A.F. Cheng[1], J.M. Moore[3], W.B. McKinnon[11], W. M. Grundy[12], J.J. Kavelaars[13], I.R. Linscott[14], W. Lyra[15], B.L. Lewis[16,17], D.T. Britt[18], J.R. Spencer[2], C.B. Olkin[2], R.L. McNutt[1], H.A. Elliott[5,19], N. Dello-Russo[1], J.K. Steckloff[20,21], M. Neveu[22,23], and O. Mousis[24]

[1] Space Exploration Sector, Johns Hopkins University Applied Physics Laboratory, 11100 Johns Hopkins Rd, Laurel, MD USA 20723 carey.lisse@jhuapl.edu, hal.weaver@jhuapl.edu, ralph.mcnutt@jhuapl.edu, andy.cheng@jhuapl.edu, neil.dello.russo@jhuapl.edu

[2] Southwest Research Institute, Boulder, CO, USA 80302 alan@boulder.swri.edu, layoung@boulder.swri.edu, joel@boulder.swri.edu, colkin@boulder.swri.edu, spencer@boulder.swri.edu

[3] Astrophysics Branch, Space Sciences Division, NASA/Ames Research Center, Moffett Field, CA, USA 94035 dale.p.cruikshank@nasa.gov, scott.a.sandford@nasa.gov, orkan.m.umurhan@nasa.gov

[4] Université Grenoble Alpes, CNRS, CNES, Institut de Planétologie et Astrophysique de Grenoble, Grenoble, France bernard.schmitt@univ-grenoble-alpes.fr

[5] Solar System Exploration Research Virtual Institute, NASA/Ames Research Center, Moffett Field, CA,USA 94035 yvonne.pendleton@nasa.gov, jeff.moore.mail@gmail.com

[6] Astrophysics and Space Sciences Section, Jet Propulsion Laboratory/Caltech, Pasadena CA 91109, USA jkeane@caltech.edu

[7] Southwest Research Institute, San Antonio, TX, USA 28510 randy.gladstone@swri.org, helliott@swri.edu

[8] Dept. of Earth, Atmospheric, and Planetary Sciences, Massachusetts Institute of Technology, Cambridge, MA, USA 02139 rpb@mit.edu, aearle@mit.edu

[9] Laboratory for Atmospheric & Space Physics, University of Colorado, Boulder, CO, USA 80303 mihaly.horanyi@lasp.colorado.edu

[10] Birkbeck, University of London, WC12 7HX, London, UK m.elmaarry@bbk.ac.uk

[11] Department of Earth and Planetary Sciences and McDonnell Center for Space Sciences, One Brookings Drive, Washington University, St. Louis, MO 63130 mckinnon@wustl.edu

[12] Lowell Observatory, 1400 West Mars Hill Road, Flagstaff, AZ 86001 W.Grundy@lowell.edu

[13] NRC Herzberg Inst of Astrophysics, 5071 W Saanich Road, Victoria BC V9E 2E7 BC, Canada JJ.Kavelaars@nrc-cnrc.gc.ca

[14] Hansen Experimental Physics Laboratory, Stanford University, Stanford, CA 94305-9515 linscott@stanford.edu

[15] Dept. of Astronomy, New Mexico State University, PO BOX 30001, MSC 4500, Las Cruces, NM 88003-8001 wlyra@nmsu.edu

[16] Department of Astronomy, Columbia University, 550 W. 120th St., New York, NY 10027 bll2124@columbia.edu

[17] Division of Astronomy and Astrophysics, University of California, Los Angeles, 475 Portola Plaza, Los Angeles, CA 90025

[18] Department of Physics, University of Central Florida, Orlando, FL 32816 dbritt@ucf.edu

[19] Physics and Astronomy Department, University of Texas at San Antonio, San Antonio, TX 78249, USA

[20] Planetary Science Institute, Tucson, AZ 85719 jordan@psi.edu

[21] Department of Aerospace Engineering and Engineering Mechanics, University of Texas at Austin, Austin, TX, 78712

[22] Department of Astronomy, University of Maryland College Park, College Park, MD 20742 marc.f.neveu@nasa.gov

[23] NASA/Goddard Space Flight Center, Planetary Environments Laboratory, Greenbelt, MD 20771

[24] Aix-Marseille Université, CNRS, CNES, LAM, Marseille, France olivier.mousis@lam.fr

34 Pages, 5 Figures, 2 SOM Tables

## Abstract

In this paper we discuss in a thermodynamic, geologically empirical way the long-term nature of the stable majority ices that could be present in Kuiper Belt object (KBO) 2014 $MU_{69}$ (also called Arrokoth; hereafter "$MU_{69}$") after its 4.6 Gyr residence in the Edgeworth-Kuiper belt (EKB) as a cold classical object. We compare the upper bounds for the gas production rate ($\sim 10^{24}$ molecules/sec) measured by the New Horizons (NH) spacecraft flyby on 01 Jan 2019 to estimates for the outgassing flux rates from a suite of common cometary and KBO ices at the average $\sim 40K$ sunlit surface temperature of $MU_{69}$, but do not find the upper limit very constraining except for the most volatile of species (e.g. CO, $N_2$, $CH_4$). More constraining is the stability versus sublimation into vacuum requirement over Myr to Gyr, and from this we find only 3 common ices that are truly refractory: HCN, $CH_3OH$, and $H_2O$ (in order of increasing stability), while $NH_3$ and $H_2CO$ ices are marginally stable and may be removed by any positive temperature excursions in the EKB, as produced every $10^8 - 10^9$ yrs by nearby supernovae and passing O/B stars. To date the NH team has reported the presence of abundant $CH_3OH$ and $H_2O$ on $MU_{69}$'s surface (Stern *et al.* 2019, Grundy *et al.* 2020). $NH_3$ has been searched for, but not found. We predict that future absorption feature detections, if any are ever derived from higher signal-to-noise ratio spectra, will be due to an HCN or poly-$H_2CO$ based species. Consideration of the conditions present in the EKB region during the formation era of $MU_{69}$ lead us to state that it is highly likely that it "formed in the dark", in an optically thick mid-plane, unable to see the nascent, variable, highly luminous Young Stellar Object (YSO)/TTauri Sun, and that KBOs contain HCN and $CH_3OH$ ice phases in addition to the $H_2O$ ice phases found in their short period (SP) comet descendants. Finally, when we apply our ice thermal stability analysis to bodies/populations related to $MU_{69}$, we find that methanol ice, unlike the hypervolatile ices, should be ubiquitous in the outer solar system; that if Pluto isn't a fully differentiated body, then it must have gained its hypervolatile ices from proto-planetary disk (PPD) sources in the first few Myr of the solar system's existence; and that hypervolatile rich, highly primordial comet C/2016 R2 was placed onto an Oort Cloud orbit on a similar timescale.

# 1.    Introduction.

On 01 Jan 2019, the New Horizons spacecraft visited for the first time ever a small, pristine cold classical Kuiper Belt object (KBO), Arrokoth (2014 $MU_{69}$; hereafter $MU_{69}$). Resolved multiband photometric imaging and near-infrared (NIR) spectroscopy were obtained of $MU_{69}$'s surface, producing a picture of a dark, highly reddened object containing solid water ice, methanol, and tholins (Stern *et al.* 2019) and no detectable gas in any surrounding coma.

Because its orbit is nearly circular and uncoupled to the solar system's planets, $MU_{69}$ is thought to have formed in place 4.6 Gyrs ago at ~45 AU, and could contain one of the most primitive collections of early solar system material studied to date. Naively, one might think that its current low ambient dayside average temperature of ~42 K should have kept most icy species in "deep freeze", leaving them unreacted and unchanged since their formation. The ice species of interest include those detected on planetary satellites, Centaurs, Pluto, and KBOs in the outer Solar System like $H_2O$, $CH_4$, $N_2$, CO, $CO_2$, $CH_3OH$ (methanol), HCN (hydrogen cyanide), $NH_3 \bullet nH_2O$ (ammonia hydrate), and $C_2H_6$ (ethane); and those detected in cometary comae as the products of nuclear mass loss, such as $C_2H_2$, $C_3H_8$, $SO_2$, and $O_2$ (Bieler *et al.* 2015, Mall *et al.* 2016).

However, while $CH_3OH$ was found from two well-defined absorption bands at 2.27 and 2.34 μm by the New Horizons Linear Etalon Imaging Spectral Array LEISA (LEISA) 1.25 - 2.5 μm mapping spectrometer ($\lambda/\Delta\lambda$ = 240 and $MU_{69}$ spatial resolution ~1.8 km $px^{-1}$ for a, 33 x 17 km object; Reuter *et al.* 2008, Stern *et al.* 2019, Spencer *et al.* 2019; Figure 1), $H_2O$ signatures were only weakly present if at all (as seen by a broad $H_2O$ absorption band centered at 2.0 μm; Lisse et al. 2017, Grundy et al. 2020). As for $CH_4$, none of the multiple prominent absorption bands in the LEISA spectral range were seen. Nor were the extremely "hypervolatile" species CO and $N_2$ detected. Among the weakly absorbing species, the important candidate molecule HCN has three small absorption bands in the LEISA range ($2\nu_3$ at 2.432 μm, ($\nu_1 + \nu_3$) at 1.866 μm, and $2\nu_1$ at 1.535 μm; Quirico *et al.* 1999), that could not be discerned within the noise pattern of the spectra obtained. Similarly, $CO_2$ was not detected; its principal signature in the LEISA spectral range consists of three narrow and relatively weak bands near 2.0 μm which, if present, could not be discerned at the noise level achieved in the data. ((The returned spectra, while revolutionary for such a small KBO, were also at the same time limited by the finite pixel scale and rapid scan pattern of the fast New Horizons

flyby.)

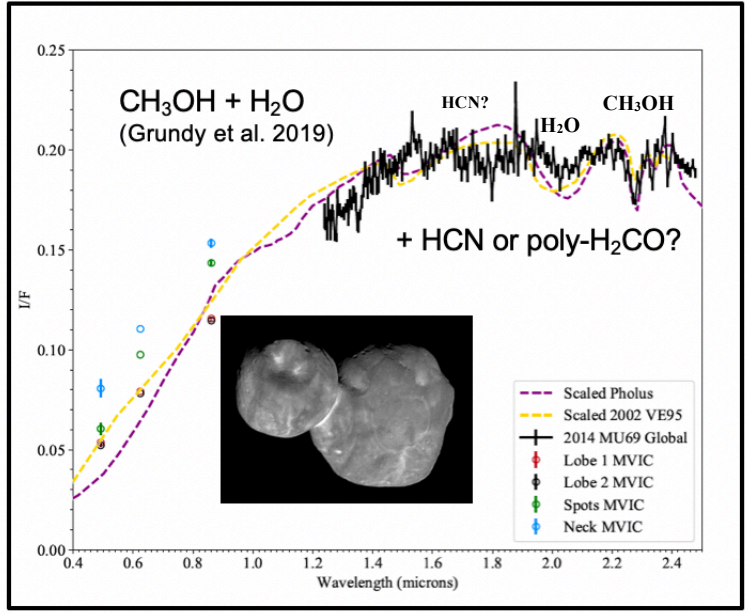

Figure 1 – Ices found on Arrokoth (top) and Pluto (bottom) by New Horizons' Ralph/LEISA instrument suite to create spectrophotometric maps of the bodies. MU$_{69}$'s surface is dominated by strong, nearly uniform signatures of CH$_3$OH and a reddish-orange tholin, and weak signatures of H$_2$O. In stark contrast, Pluto's surface is dominated by CH$_4$, N$_2$, CO, and reddish "Pluto tholin".

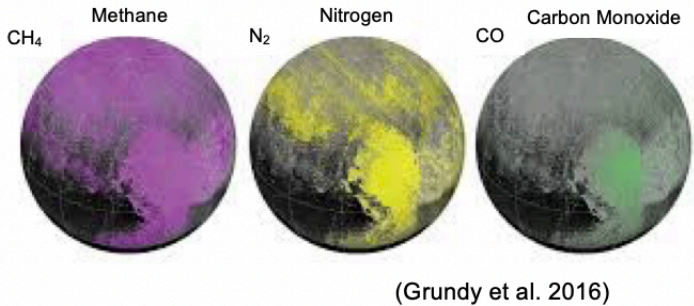

(Grundy et al. 2016)

In this paper, we seek to provide a thermal stability interpretation for the observed composition of MU$_{69}$'s surface. To this end, we utilize thermodynamical studies of laboratory analogue ice vapor pressures and find that almost all of the primitive ices known to be in outer solar system icy bodies should have been removed long ago from a nigh-gravitationless MU$_{69}$. Our analysis suggests that only the high boiling temperature (hereafter "refractory"), hydrogen-bonded ice species H$_2$O, poly-H$_2$CO, CH$_3$OH, and HCN, as well as minority impurities in refractory-ice phases, should be present if accreted by MU$_{69}$ 4.6 Gyr ago. Our analysis also reproduces the observed mix of solid, stable ices on Pluto's surface and the N$_2$/CO/CH$_4$ found in its atmosphere (Grundy *et al.* 2016; Figure 1).

We interpret the surface composition to also reflect the bulk composition of MU$_{69}$ and other distant small bodies with a similar dynamical history. Although in principle the New Horizons LEISA spectra and MVIC 4-color photometric maps (Stern *et al.* 2019, Grundy *et al.* 2020) characterize only the first few optically active microns of MU$_{69}$ surface, they show a remarkably homogenous surface throughout terrain that varies kilometers in relief across hills, valleys, rills, and crater floors on the surface of a markedly flattened object only a few kilometers thick. Further, when we look by extension to other similar distant EKB bodies in the same thermal zone, like Plutino (55638) 2002 VE95 and distant Centaur 5145 Pholus, we see similar compositional spectra (Fig. 1). Yet when we follow these objects in, we see them become active and lose their abundant methanol, and convert to the water ice dominated inner centaurs and short period comets. None of these objects show any evidence for abundant (vs water) hypervolatiles such as the CO, $N_2$, and $CH_4$ seen on Pluto and the larger KBOs. There has, however been one recent small icy body counterexample that does appear to be just as expected for a small, primordial hypervolatile-rich object - Comet C/2016 R2, known for its large outgassing of CO and $N_2$ but no water. As we discuss below, this makes sense if R2 has retained its hypervolatile ices to the present day, and is only now subliming some of them while keeping the rest of its volatile ices ultra-cold, stable, and rock hard via sublimative cooling.

Section 1 of this paper is intended to present the known observational results for ices in MU69 from the New Horizons flyby. Each alone is moderately useful, but the combination of surface spectral mapping, UV coma non-detection, and rotationally resolved body imaging provide powerful constraints on the nature of the body. In Section 2 we present the details of our thermodynamic calculations and the resulting saturation vapor pressure versus temperature ($P_{sat}$ vs T) curves.[1] In this Section we also calculate the upper limits for the surface area of any currently exposed sublimating ices presented by the non-detection of a gas coma around $MU_{69}$ by New Horizons.

---

[1] Section 2 and the SOM are important for explicitly writing down for both the KBO and the Comet communities' use the temperature dependence of ice vapor pressure for the dominant ice species expected in outer solar system icy bodies. In doing this, we employ the measured heats of vaporization for the pure ice species which are readily obtainable from terrestrial laboratory measurements. Real KBO ices are likely to be some sort of mixture of pure and compounded (or alloyed, or mixed; in this paper we use the terms interchangeably) ices, due to the chaotic messiness of planetesimal formation in the early solar system, with a PPD likely dominated by number by simple $H_2O$, CO, and $N_2$ molecules plus heavier species created via heterogenous catalysis on grain surfaces (e.g. Nuth *et al.* 2008). Add to this the vagaries of the temperature and density structure of the convecting PPD, the effects of protoplanet disk stirring, and the behavior of stochastic solar FU Ori/EX Lupi accretion events and XUV stellar outbursts over time, and it is quite difficult to state with any certainty the exact nature (pure, mixed, alloyed, or crystalline) of the ices first incorporated into proto-KBO bodies. Until we have a cryogenic core sample of these ices in a terrestrial laboratory for measurement, we will not know which of a nearly infinite possible mixed phases we should measure; fortunately the properties of many possible combinations of mixed ices have been studied by the Ames group in the 90's [c.f. Sandford *et al.* 1988, 1990, 1993], and they grossly tend to follow the dominant ice species' behavior, which we study here.

Using these laboratory measured heat of sublimation values, in Section 3 we then show that if low boiling point temperature molecular ice species like $N_2$, CO, or $CH_4$ (hereafter "hypervolatile ices") were *ever* incorporated into MU69 as majority ice phases, then they are not only not there today, but they shouldn't have been there after the first few Myr of its existence. By contrast, high boiling point temperature hydrogen bonded ice species like $H_2O$, $CH_3OH$, or HCN (hereafter "refractory ices") incorporated as majority phases into MU69 4.56 Gyrs ago should still be there today, stable as rock.

The final sections of the paper are interpretive of the facts established in the first two sections. In Sections 4 we show analytically how thermally driven icy body evolution can explain the NH findings for MU69. In Section 5 we extend our analysis via logical arguments to predictions bearing further verification concerning the compositional connections between KBOs, Centaurs, and Comets, the expected abundant refractory ices composing the Pluto system's bodies, and the curious hypervolatile rich counter-case of Comet 2016 R2.[2]

## 2.  Thermodynamical Background.

Investigations on the subject of the stability of EKB ices have been conducted intermittently in the literature over the last few decades. We summarize the major efforts here in order to provide the reader with a useful summary of previous thought and likely avenues for future research and progress. (This work, for example, was inspired by our 2019 understanding that updated laboratory cryogenic heat of sublimation measurements produced by KBO astronomers have not yet been applied to the small icy body sublimation models produced by cometary astronomers or the hot core ice models of astrophysicists.)

---

[2] Caveat: By no means is this treatment meant to replace detailed time-dependent modeling of a 3-dimensional representation of KBO, which is why we do not delve deeply in this paper into the details of the first ~1 Myrs of Arrokoth's life - which must have been very interesting and variable, although transient. The purpose of this paper is to set up the problem of the possible icy constituents of KBO MU69 and Pluto as seen today after 4.56 Gyr of thermal evolution. I.e., to state what we know for sure from simple thermodynamic and geophysical considerations, while avoiding the uncertainties in important parameters like the runs of interior density, thermal heat capacity, and thermal conductivity through the body that produce a variety of outcomes in more detailed models. We hope that this work and its given material abundance restrictions inspires future detailed modeling work that will fill in more of the picture, especially the short-term time dependent behavior of KBO interiors over the first few Myr, by studying the extent of modeling phase space in the manner of Merk *et al.* 2006 and Prialnik 2009.

Starting in the late 1980s, astronomical infrared spectral studies investigating the composition of icy molecular cloud cores, the precursors to solar systems and their icy planetesimals, detected absorption features indicative of ices containing $H_2O$, CO, $CO_2$, $CH_4$, $H_2CO$, $NH_3$, and $CH_3OH$ (Tielens *et al.* 1989, Allamandola *et al.* 1992, Lacy *et al.* 1998). The spectral and physical properties of these ices, including their sublimation and condensation behaviors, were studied in the laboratory (e.g., Sandford & Allamandola 1988, 1990, 1993a, 1993b; Schutte *et al.* 1993a,b) to investigate the possible makeup of these clouds, and the plausibility that the laboratory ice analogues could be present in the molecular cloud at the ambient temperatures estimated from their spectroscopy. This work produced surface binding energy estimates for the various ices, useful for producing stability ages for their solid, icy phases assuming the relative "sticking probabilities" of molecules as a function of temperature.

For a gaseous species in chemical equilibrium with its solid phase the dependence of its saturation vapor pressure is given by the Clausius-Clapeyron thermodynamic equation:

(1) $$1/P \, dP_{sat}/dT = \mu H_{subl}/R_g T^2$$

with solution $P_{sat} = C*exp^{(-\mu H subl/R_g T)}$. Here, $P_{sat}$ is the saturation vapor pressure above a patch of ice, T is the ice temperature, $\mu$ is the mass of 1 mole of gas, $H_{subl}$ denotes the enthalpy (i.e., latent heat) of sublimation per kg, and $R_g = N_{Avogadro}*k_{Boltzmann}$ is the ideal gas constant. Assuming that $H_{subl}$ is independent of temperature, Prialnik *et al.* (2004), following up on earlier laboratory work in forming and evolving analogue icy cometary materials (Yamamoto 1983; Bar-Nun *et al.* 1987; Prialnik *et al.* 1987, 1990; Schmitt 1989, 1992), used this relation together with the empirically tabulated $P_{sat}(T) = A * exp^{(-B/T)}$ fits known since the 1920s to obtain and tabulate its exponential coefficient in terms of the latent heat (SOM Table 1). The form of these $P_{sat}$ curves is such that they all have a universally similar structure with rapidly rising pressures and "knees" controlled by the value of $(H_{subl}/k_{Boltzmann})$, i.e. the heat of sublimation expressed in Kelvin. Figure 2 shows examples of these curves, where it can be seen that ices which sublime easily (like hypervolatile molecular solids $N_2$, CO, and $CH_4$ with only van der Waals molecule-molecule interactions) have small values of $H_{subl}/k_{Boltzmann}$, on the order of $10^2$ K, while refractory ices (like the hydrogen-bonded species $H_2O$, $H_2CO$, $CH_3OH$, $HCN$, and $NH_3$ with relatively strong inter-molecular bonds) have $H_{subl}/k_{Boltzmann}$ values in the many-thousands (e.g. Sandford & Allamandola 1993a, 1993b).

The flux rate at which an icy cometary material loses molecules into a surrounding medium follows from gas kinetic theory (Prialnik *et al.* 2004) as

(2)     $Z$ (molecules/m²/s) = $\rho_{gas}*v_{gas} = (P_{sat}-P_{ambient})/kT*v_{thermal, z+} = (P_{sat}-P_{ambient})/(2\pi kTm)^{0.5}$

where $Z$ is the mean loss rate per unit normal surface area, the ideal gas law has been used to express the gas number density $\rho_{gas} = (N/V) = P/kT$, $v_{thermal,z+} = (kT/2\pi m)^{0.5}$ is the mean Maxwell-Boltzmann velocity at temperature T for gas molecules leaving perpendicular to the ice surface, m is the molecular gas mass, and k is Boltzmann's constant, respectively.

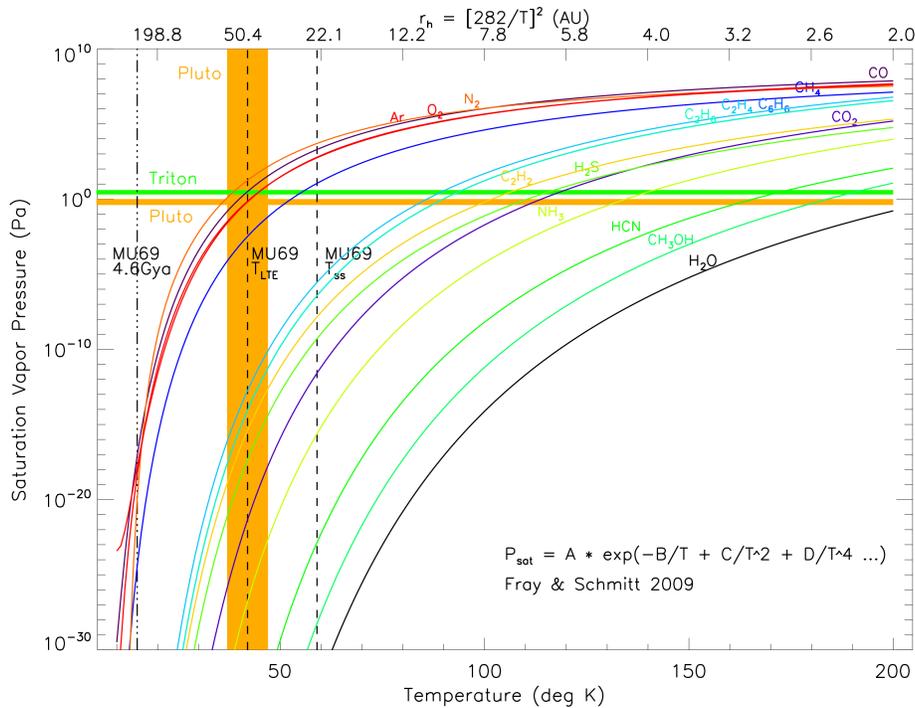

Figure 2 – Saturation pressure $P_{sat}$ as a function of temperature for 12 common species found in small solar system body ices. The range of saturation pressure curve behavior ranges from hypervolatile ($N_2$, CO, $O_2$, Ar, $CH_4$) at upper left to highly refractory (HCN, $CH_3OH$, $H_2O$) at lower right. In the mid-range are a host of medium-volatile hydrocarbon species. Also shown is the range of temperatures (37-47 °K) and pressures (~10 μbar) on Pluto's surface determined by New Horizons (intersection of horizontal and vertical gold bars). By comparing to our model $P_{sat}$ curves, it can be seen that only the hypervolatile ices will be in the gas phase on Pluto, and that in some regions they will be stable solids.

Once the proper local equilibrium temperature T is calculated (see Section 4.1), two important special cases of Eqn. 2 occur:

(1) When $P_{sat} < P_{ambient}$, as for a body with a stable atmosphere, no net molecules are actively emitted into the medium and the ice is thermally stable and remains condensed. This is the case studied in the seminal paper of Schaller & Brown (2007) predicting which of the largest KBOs could retain atmospheres and thus finite surface pressures, and which ices would be stable given the KBOs ambient temperature and pressure (Schaller & Brown 2007, Brown *et al.* 2011). We can use this approach to predict the ices which should occur on Pluto's surface and in its atmosphere given the ~10 μbar of ambient surface pressure (as measured by New Horizons, Stern *et al.* 2015, Gladstone *et al.* 2017).

(2) When $P_{ambient} \approx 0$, in which case the ice is exposed to vacuum, and the derived sublimative flux rates can be compared to the ~$10^{15}/cm^2$ molar surface densities of ices[3] to determine how long it takes for a patch of exposed ice to fully sublimate away. This case is analogous to the residence lifetime arguments of Langmuir (1916), Frenkel (1924), and Sandford & Allamandola (1993b), and can be used to predict the lifetime (versus thermal sublimation) of exposed ices on comet surfaces and in interstellar and interplanetary dust grains. These lifetimes are upper limits, as other removal processes, such as photon irradiation, stellar wind particle sputtering, and micrometeorite impact gardening can also remove mass from exposed ice.

To confirm which regime applies to MU$_{69}$ and Pluto, we use the Catling & Zahnle (2009, 2017) airless body condition. A body can retain a stable atmosphere when $V_{thermal} = 0.8$ km/sec $*(T/300K)^{0.5} << V_{escape} = (2GM_{KBO}/R_{KBO})^{0.5} = (8\pi G\rho R_{KBO}^2/3)^{0.5}$ ~11 km/sec $* (R_{KBO}/6371$ km$)^{2/3}$ $*(\rho/5.2 gcm^{-3})^{0.5}$, while the $P_{ambient} \approx 0$ condition is pertinent when the opposite is true for a body, $V_{thermal} >> V_{escape}$. Evaluating, we have for Pluto that $V_{thermal}$ ~250-320 m/sec $< V_{escape}$ ~1200 m/s, while for MU$_{69}$, $V_{thermal}$ ~290 m/s $>> V_{escape} \sim 4$ m/s.

---

[3] Molar surface density ~ $(\rho/MW * N_{Avogadro})^{0.67}$ ; for water ($H_2O$) with bulk density $\rho = 1 g/cm^3$ and MW = 18 g/mole, this is $1.0x10^{15}/cm^2 => 3.2x10^7$ molecules/cm, or $3.1x10^{-8}$ cm/molecule = 3.1 Å/molecule. For methanol ($CH_3OH$) with bulk density $\rho = 0.8 g/cm^3$ and MW = 32 g/mole, its molar surface density is $(0.8/32 * 6x10^{23})^{0.67} = 6.1x10^{14}/cm^2$, and the intramolecular spacing is $10^8 Å/cm/(6.1x10^{14}/cm^2)^{0.5} = 4.0$ Å/molecule. For HCN with bulk density $\rho = 0.7$ g/cm$^3$ and MW = 27 g/mole, its molar surface density is $(0.7/27 * 6x10^{23})^{0.67} = 6.2x10^{14}/cm^2$, and the intramolecular spacing is $10^8 Å/cm/(6.2x10^{14}/cm^2)^{0.5} = 4.0$ Å/molecule.

Fray & Schmitt (2009) revisited the problem of the temperature dependence of $P_{sat}$, going beyond the simple assumption of $H_{subl}$ = Constant and allowing for temperature variation of the heat of sublimation. After conducting an exhaustive literature search for relevant ice data, they reported results for more than 20 species in polynomial form, as $P_{sat} = A_0 + A_1/T + A_2*T^2 + A_3/T^3, +A_4/T^4 + A_5/T^5$ (reproduced here in SOM Table 2) greatly refining the accuracy of the $P_{sat}$ curves, especially at low values of $(R_gT/H_{subl})$ where $P_{sat}$ changes most rapidly. Since we are most interested in the behavior of ices in small outer solar system bodies at low temperatures (T = 10 to 90 K), we have adopted the more refined constants published by Fray & Schmitt (2009) (SOM Table 2) in calculating the values of $P_{sat}$ for this work.

## 3. Results.

Figure 2 shows the resulting $P_{sat}$ saturation pressure curves we have calculated from the Fray & Schmitt data. Some immediate findings can be made from them:

(1) As a check of our calculations, the $P_{sat}$ values derived from Prialnik *et al.* 2004's $H_{subl}$ = constant approximation and the Fray & Schmitt 2009 polynomial fits were compared, and found to match well at high temperatures T > $(H_{subl}/k)$ where the temperature dependence of $H_{subl}$ match is small.

(2) There are hypervolatile ($N_2$, CO, $CH_4$), mid-volatile ($C_2H_2$, $C_2H_4$, $C_2H_6$, $C_6H_6$, $H_2S$, $CO_2$, $NH_3$), and 'refractory' (HCN, $CH_3OH$, $H_2O$) ices in the list, with a broad range of outgassing rates at any given temperature.

(3) All but the most hypervolatile ($N_2$, CO, $O_2$, Ar, $CH_4$) species should be ***stable as rock-solid ices*** on the surface of Pluto today, with its ambient pressure of ~10 μbar = 0.1 Pa. The hypervolatile species should be ***metastable*** depending on the local surface temperature (ranging from 37K in Sputnik Planitia to 45 K in Cthulhu Macula; Earle *et al.* 2017) and pressure (ranging from 11 μbar in lowland Sputnik Planitia to ~8 μbar in the mountain highlands; Schmitt *et al.* 2017, Moore *et al.* 2018, Young *et al.* 2018, Bertrand *et al.* 2019). This is consistent with New

Horizons' determination of the ($N_2$, CO, $CH_4$) surface patterning and atmospheric makeup of Pluto (Stern *et al.* 2015, Gladstone *et al.* 2016, Young *et al.* 2018).

(4) For similar reasons, $MU_{69}$, with its P $\approx$ 0 surface pressure, should be unable to retain the hypervolatile ($N_2$, CO, $O_2$, Ar, $CH_4$) species in the form of pure ices. At the local thermodynamic equilibrium temperature of ~42 K in its stable cold classical orbit, for an object with emissivity = 0.90 (Stern *et al.* 2019, Buratti *et al.* 2019, Verbiscer *et al.* 2019), these hypervolatile ices have high vapor pressures (as demonstrated by their presence in Pluto's atmosphere, above Pluto's even colder surface; Fig. 2). In fact, these species should never have been condensed out of the gas phase in any appreciable amounts in the vacuum of space at current $MU_{69}$ temperatures. For these gases to condense originally as dominant ice phases, $MU_{69}$ would have had to have formed in a much colder locale, as could be created by a proto-Edgeworth-Kuiper Belt heavily shrouded from the proto-Sun by an optically thick proto-planetary disk mid-plane. In this case, the lowest local temperatures would have been that of molecular clouds in the galactic interstellar medium (ISM) radiation field, or about 15 K. (Observations of currently extant exo-PPDs (e.g., Anderl *et al.* 2016, Öberg *et al.* 2017, Loomis *et al.* 2020) and models of the solar nebula and PPDs around Sun-like stars (e.g., Lesniak & Desch 2011; Krijt *et al.* 2018, 2020; Mousis *et al.* 2019) have mid-plane temperatures at 45 AU from the Sun at T ~ 15 to 25 K within the first few Myr of the solar system's existence.) From Fig. 2, temperatures at this level can allow appreciable hypervolatile ice accumulation, and the detection of hypervolatile ices in interstellar dense clouds attests to this occurrence (Tielens *et al.* 1989, Allamandola *et al.* 1992, Lacy *et al.* 1998). But it is also during the first few Myr of a KBO's existence when short-lived radioactive nuclides like $Al^{26}$ can cause substantial warming in the heart of a KBO (Choi *et al.* 2001, Prialnik 2002, Prialnik *et al.* 2004), making the condensation of hypervolatiles problematic.

In the next section, we show that if bulk majority phase hypervolatile ices were initially incorporated into $MU_{69}$, then subsequent short-lived radioactive driven evolution would relegate them to near-surface regions, and that these regions will then become depleted within ~1 Myr of the present-day surface temperatures being established. We go on to state that the hypervolatile ice species CO and $CH_4$ we see outgassing from icy bodies today must have been stabilized over time by residing as a minority component in a non-pure ice whose thermal stability was controlled by a more refractory ice like $CH_3OH$ or $H_2O$.

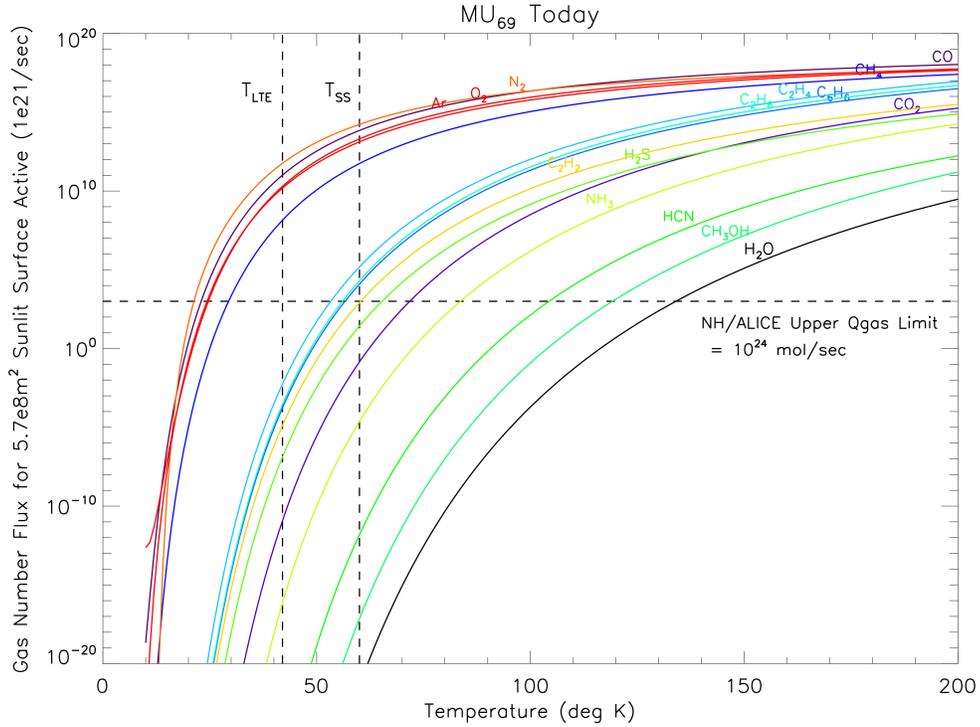

Figure 3 – Plot of total $Q_{gas}$, the total production rate of gas molecules for a given ice species from $MU_{69}$, vs ambient temperature, assuming the ice species is uniformly distributed across $MU_{69}$'s sunlit side. The non-detection for gas production produced using the NH/ALICE spectrometer's measurements (of $H_2$) is shown by the dashed line. If the surface of $MU_{69}$ contained appreciable $N_2$, CO, $O_2$, Ar, or $CH_4$ icy material at ~42K in thermal contact with the surface, we would have detected its gas molecule emission rate above the limiting sensitivity of the ALICE measurements.

Observational support for the lack of hypervolatile ices (except as minority impurities in dominantly refractory ice phases) in $MU_{69}$ is shown in Figure 3. Here we have compared the 3-sigma upper limit on the gas production rate, $Q_{gas}$, supporting the gravitationally unbound atmosphere, or coma, surrounding $MU_{69}$ as determined by NH/ALICE UV spectrometer solar Lyman-alpha airglow measurements during the 01 Jan 2019 flyby to our calculated net outgassing rate for ices mixed uniformly on the surface of a $T_{LTE}$ = 42 K object of $MU_{69}$'s measured surface area. The published upper limit for $Q_{gas}$ is ~ $10^{24}$ molecules/sec (Stern *et al.* 2019), a relatively high rate of gas release at 45 au when compared to other actively subliming solar system bodies like comets at ~1 au (Lisse 2002, Bockelee-Morvan *et al.* 2004). However, this value is sensitive enough to rule out uniform layers of hypervolatile $CH_4$ ice, and, by extension, ($N_2$, CO, $O_2$, and Ar), on direct measurement grounds, as these species would produce gas flux rates far above the upper limit rate (dashed line) in Fig. 3.

Further, if we note that MU69 had a special orientation geometry during the NH flyby such that one-half of its highly flattened surface was pointed nearly sunward while the other half was in near-total shadow, so that the sunlit side temperature was closer to $T_{ss} = 59K$, then we see that even the moderately volatile organic ice species $C_2H_2$, $C_2H_4$, $C_2H_6$, and $C_6H_6$ are ruled out.

## 4.     Discussion & Analysis.

More stringent constraints can be placed on ice stability if we consider not only their current sublimation loss rate, but their loss rate over the age of the Solar System and the age of MU69. Using the outflux rates of Fig. 3, as a most-conservative case we can assume their stability over time in the EKB during their Gyrs-long post-aggregation phase, and see how long it takes a molar surface density's worth[1] ($\sim 10^{15}/cm^2 = 10^{19}/m^2$) of material to be removed via thermal sublimation.

4.1     Negligible Sublimative Cooling After $\sim 1$ Myr. We can safely make an approximate calculation using simple thermodynamic and energy balance arguments on Gyr timescales for the following reason. The energy balance equation (neglecting any endogenic heat flux) for heating of a unit slab of KBO material (1m wide by 1 m long by dz thick) of Bond albedo A and effective thermal emissivity ε oriented at angle ξ at $r_h$ distance from a 1 $L_{sun}$ luminosity star is :

(3)     $(1-A)L_{sun} \cos\xi / 4\pi r_h^2 = \varepsilon\sigma T^4 + \Sigma_i Z_i \Delta H_{subl,i} * (10^{-3} M.W.(g)/N_{Avogadro})_i$

where the first term is any input insolation, the second is the re-emitted thermal radiation from the object, and the third is the cooling derived from sublimation of the object's volatile species. We look at the relative magnitudes of these terms by calculating the maximal possible rate of sublimative cooling vs. temperature possible for an icy species, using the gas production rate ($Q_{gas}$, in molecules/m²/s) curves of Fig. 3 multiplied by the heat of sublimation ($H_{subl,i}$, in J/molecule sublimed) for a species. Figure 4 shows the result, from which it can be seen that sublimative cooling is a negligible term for an object like MU69 in the Edgeworth-Kuiper Belt compared to the $\sim 0.7$ W/m² received from the Sun, except in the case of the hypervolatile species (CO, $N_2$, $CH_4$,

Ar, and $O_2$). This result is also consistent with other ice sublimation modeling work, like that of Steckloff *et al.* (2015, 2020).

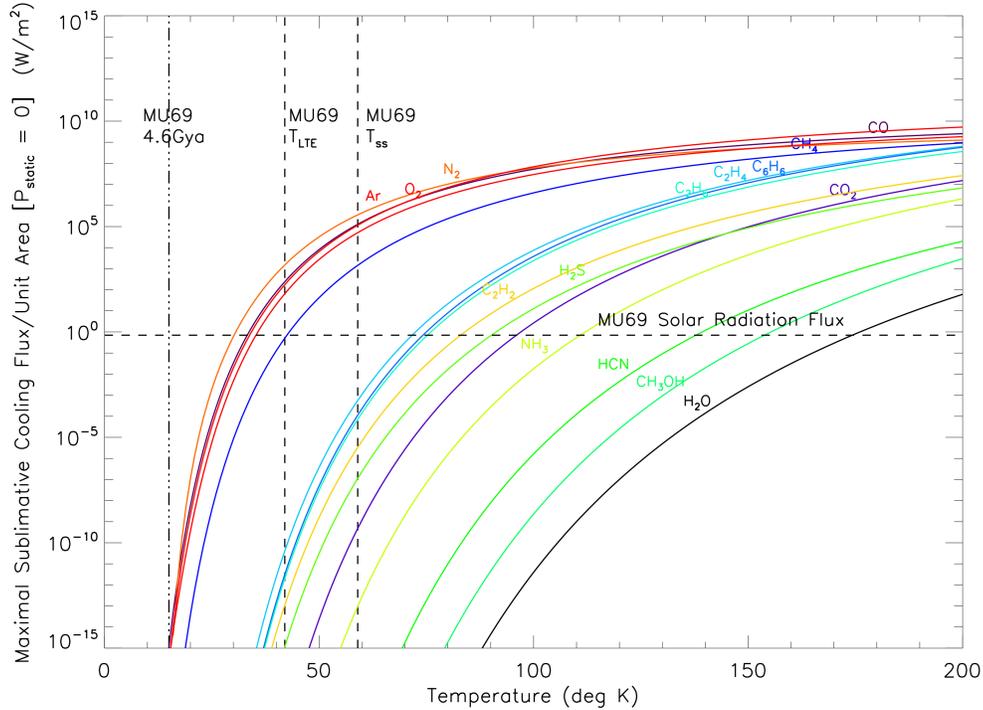

Figure 4 – Plot of the maximal possible sublimative cooling power per unit surface area for a given ice species vs. ambient temperature, assuming the ice species is uniformly distributed across $MU_{69}$'s sunlit side. Only for the hypervolatile species ($N_2$, CO, $O_2$, Ar, or $CH_4$) does the sublimative power achieve or exceed the input insolation power.

We have shown in Section 3 that there are no dominantly hypervolatile ice phases in abundance today on the surface of $MU_{69}$, and argued that it is unlikely that any majority hypervolatile ices ever did condense into the KBO unless it formed in a cold, optically thick mid-plane region of the PPD. In this eventuality, the effects of sublimative cooling due to hypervolatile evaporation would be short-lived. E.g., if we were to construct $MU_{69}$ entirely out of albedo = 0.16 $N_2$ or CO ice with $H_{sub}$ ~$10^{-20}$ J/molecule (Prialnik *et al.* 2004, Fray & Schmitt 2009), it would take ~$4\times10^{28}$ molecules/sec (~$10^3$ kg/sec) subliming (close to the typical activity level for an inner system SP comet near perihelion) to match the (0.7 W/m$^2$ * $6\times10^8$ m$^2$ surface area) of $MU_{69}$ total input solar power. The mass of $MU_{69}$ is about $(6\times10^8)^{3/2}$ m$^3$ * 500 kg/m$^3$ = $2\times10^{15}$ kg. Since the $10^3$ kg/sec mass loss is always "on" for a body in a near-circular, stable orbit like $MU_{69}$'s, this means an $MU_{69}$ consisting of solid $N_2$/CO would be fully sublimated in ~$10^5$ yrs, and any hypervolatiles providing substantive cooling would be quickly exhausted. (N.B. - The vast majority of small icy bodies known in the solar system are in such a hypervolatile-depleted condition. But one strong counter-example of an object temperature-controlled by apparent hypervolatile sublimation – that of comet C/2016 R2 – has been recently seen, and we discuss its case further in Section 5.5)

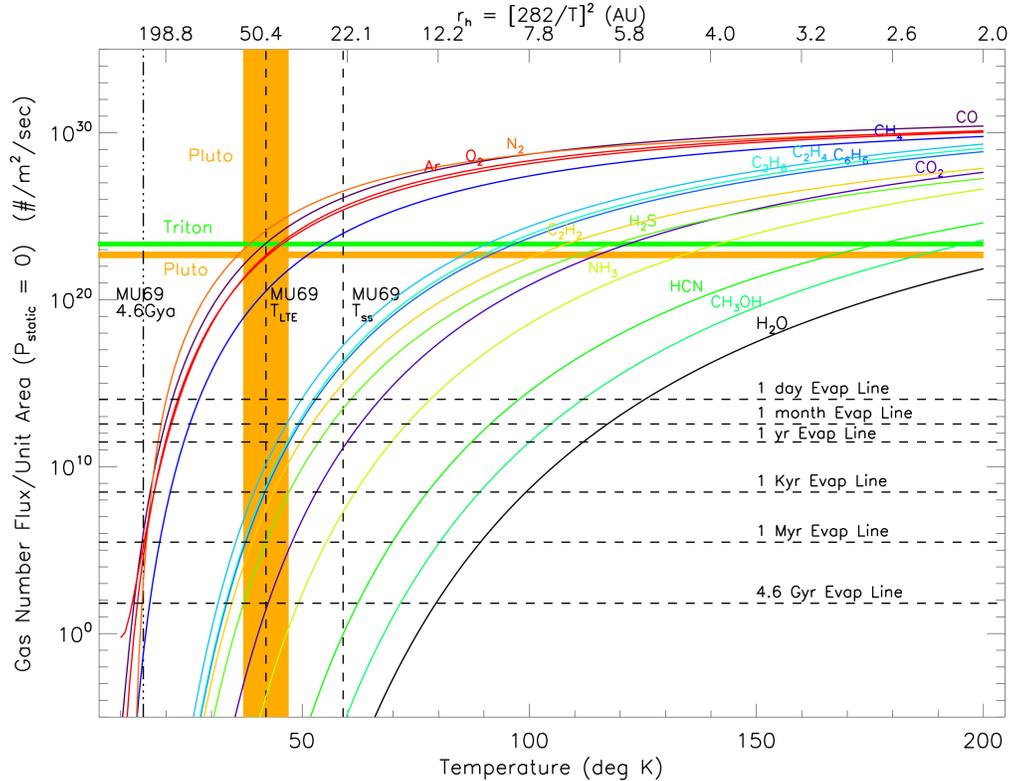

Figure 5 – Plot of $Q_{gas}$, the production rate of gas molecules for a given ice species, vs. ambient temperature, for a patch of unit surface area. This plot is useful in comparing to the molar surface density of an ice, $\sim10^{15}/cm^2$ ($=10^{19}/m^2$). For an assumed time interval over which this outgassing occurs (e.g., $10^3$ yrs ($=3\times10^{10}$ sec), $10^6$ yrs ($=3\times10^{13}$ sec), or $10^9$ yrs ($=3\times10^{16}$ sec)), the maximal $Q_{gas}$ for which any appreciable solid ice remains (rather than sublimating) can be calculated (dashed horizontal lines). While this line of analysis involves assuming constant levels of gas mass loss over large periods of time, we have assumed the lowest possible loss rates (thermal only) and the method produces much more stringent limits on the possible presence of any ices. As a result, it immediately shows that only the highly refractory hydrogen-cross bonded ices ($NH_3$, $HCN$, $CH_3OH$, $H_2O$) can be present for a surface at today's sunlit value of T = $\sim40$ K, and that any $N_2$, CO, $O_2$, Ar, or $CH_4$ ice incorporated by $MU_{69}$ when it first formed cold (at T$\sim$15K) in the optically dense mid-plane of the solar system's proto-planetary disk (PPD) was removed within $\sim1$ Myrs of the mid-plane clearing and $MU_{69}$ warming up to $\sim40$K.

We can now go back to Eqn. 3 and ask what the effect of any heating by short-lived radioactive elements like $Al^{26}$ would be on our unit slab, and answer that they would offset the effects of any hypervolatile sublimative cooling by causing substantive heating in the first few Myr and reducing the timespan over which sublimative cooling can keep the object below Local Thermodynamic Equilibrium (LTE). The short-lived radionuclides are especially efficient at doing this deep inside the KBO through additional volume radioactive heating decay and thermal conduction terms on the left-hand side of Eqn. 3 (Prialnik *et al.* 2004, 2006).

We are thus back to safely ignoring the effects of sublimative cooling on timescales of > 1 Myr, and with this zeroth-order approximation can produce a plot like that shown in Fig. 5. The same general family-of-curves shape is apparent in Fig. 5 as for the $Q_{gas}$ production rates in Fig. 3, but we have now added dashed horizontal lines at the $Q_{gas}$ rates that would deplete molar quantities ($10^{19}/m^2$ worth) of an ice species in 1 Yr, 1 Kyr, 1 Myr, and 4.6 Gyrs.

How to interpret this for an object like MU69? Also on Fig. 5 are 3 vertical lines, set at the T ~15 K of deep galactic ISM space; at the $T_{LTE}$ value of 42 K; and at the $T_{ss}$ value of 59 K. To $0^{th}$ order, without performing a detailed thermophysical model, once the hypervolatiles are gone, each portion of the surface will experience a temperature ranging from 15 to 59 K over an orbit, with an average $T_{LTE}$ of 42 K. The deepest interior regions, after the heat pulse from the short-lived radionuclides activity has dissipated and the insolation-driven thermal waves have stabilized, reaches a temperature half-way between the nightside and dayside temperature extremes of 15 and 59 K imposed by $MU_{69}$'s orbit and spin axis orientation, or ~37 K. Applying these temperature regimes to Fig. 5, it is seen that near the surface, the hypervolatile and mid-volatile ($C_2H_2$, $C_2H_6$, $H_2S$, etc.) ices are removed within ~1 Myr and that $CO_2$ ice is removed on the order of 4.6 Gyrs, the age of the Solar System. A stronger, mildly hydrogen-bonded ice like $NH_3$ is moderately stable, or meta-stable, versus sublimation over 4.6 Gyrs (but will be removed by additional energy inputs above solar, see next Section 4.2). Strongly hydrogen-bonded "refractory" ices (HCN, $CH_3OH$, and $H_2O$) should be thermally stable over the 4.6 Gyr age of the Solar System. For the deepest interior regions, refractory (HCN, $CH_3OH$, and $H_2O$) ices should be stable over 4.6 Gyrs, and possibly $NH_3$ and $CO_2$ ice as well.

4.2  Lower Limits to Outgassing Rate & Surface Re-processing. At this point we remind the reader that the mass loss rates plotted in Fig. 5 are lower bounding limits produced by insolative heating only. Any sort of effects that could cause increased loss – be they initial burdens of short-lived radioactive elements like $Al^{26}$ in the first $10^7$ yrs of the early Solar System; Gyrs long-term surface sputtering by photons, solar wind particles, galactic cosmic rays (GCRs), or micrometeorites; passage of the subsolar point over the surface, raising the local temperature from $T_{LTE}$ to $T_{subsolar}$, or from 42 to 59 K; or transient heating due to passing nearby O/B stars or nearby supernovae explosions every $10^8$-$10^9$ yrs (Stern 2003) – will only increase the ice loss rate, making the on-the-cusp metastable ices unstable as well. We thus expect $CO_2$ and $NH_3$, marginally stable even when neglecting these processes, to be unstable on $MU_{69}$'s surface, leaving HCN, $CH_3OH$, and $H_2O$ as the only ices to be present on the surface of $MU_{69}$ on 01 Jan 2019 during the NH flyby.

More needs to be said about this. Given the solar system abundance of early short-lived radioactive elements like $^{10}Be$, $^{26}Al$, and $^{41}Ca$ estimated from chondritic meteorites (Thrane *et al.* 2006, Castillo-Rogez 2007, Krot *et al.* 2012, Tang & Dauphas 2015), the models by Choi *et al.* 2002, Merk & Prialnik 2006, and Gounelle *et al.* 2008 argue for high central heating and material re-processing of the core of $MU_{69}$ in the first few Myr due to the decay of short-lived radionuclides like $^{26}Al$ (even melting in some cases, as the melting point of $CH_3OH$-$H_2O$ ice can be as low as 157 K at the eutectic composition of 80 mol% $CH_3OH$; Miller & Carpenter 1964). This is consistent with formation of $MU_{69}$ by the streaming instability, currently understood as the likeliest accretional scenario for dynamically cold classical EKB objects (Nesvorný *et al.* 2019), *i.e.* before dissipation of the protoplanetary gas after 5-10 Myr or ~10 $^{26}Al$ half-lives of 0.7 Myr. The surface of $MU_{69}$, however, only experienced low amounts of surface heating & re-processing due to the poor thermal conductivity of the body and the short timespan of the radioactive decay input. The presence at the surface today of $CH_3OH$, a species that (although unreactive on experimental timescales) is not thermodynamically stable in liquid water (Shock & McKinnon 1993), is an observational confirmation that any near-surface melting was likely limited in extent and in time. Observations of currently extant exo-PPDs (e.g., Anderl *et al.* 2016, Öberg *et al.* 2017, Loomis *et al.* 2020) and models of the solar nebula and PPDs around Sun-like stars (e.g., Lesniak & Desch 2011; Krijt *et al.* 2018, 2020; Mousis *et al.* 2019) have mid-plane temperatures at 45 AU from the Sun at T ~ 15 to 25 K within the first few Myr of the solar system's existence. So at its birth in an optically thick region of the solar proto-planetary disk, the surface of $MU_{69}$ was thus likely the coldest region of the body, with T ~ 15 to 25 K, capable of freezing out and maintaining hypervolatile ice phases (Figure 5).

Once the short-lived radionuclides decayed and the PPD cleared, allowing direct insolation at $MU_{69}$'s location within the first ~10 Myr (i.e., "morning arrived in the EKB"), $MU_{69}$'s subsolar surface rapidly warmed to ~50-60 K. This would have caused rapid sublimation of the surface hypervolatile ices (Figs. 4 & 5) and near-surface mass loss, which may be evidenced by the pit chains seen on $MU_{69}$'s limb (Stern *et al.* 2019, Spencer *et al.* 2020, Schenk *et al.* 2020) as well as by possible instances of scarp retreat seen on the small lobe (Moore *et al.* 2019). Other structures on $MU_{69}$, like the rumpled surface texture and the delineated joints between apparent subunits on the larger lobe, could also be due to significant subsurface mass wasting "deflation" of its initial structure (Spencer *et al.* 2020, Schenk *et al.* 2020). The local temperature would have fallen from this peak towards ~40 K as the Sun finished its gravitational condensation and commenced nuclear fusion burning, appearing on the main sequence within the first ~100 Myr.

Over the next ~4 Gyr, the Sun would have been a consistent heat source for $MU_{69}$, although the $MU_{69}$ environment at ~45 AU was likely subjected to a variety of other time-dependent energetic processes. Every $10^8$-$10^9$ yrs, the transient effects of nearby passing O/B stars and supernovae acting on $10^1$-$10^3$ yr timescales may have dramatically heated $MU_{69}$'s surface down to 10's of meters (Stern 2003), removing all but the most refractory of ices (Fig. 5). This refreshed and modified surface will also have been dosed by photons (down to μm depths), solar wind particles (down to cm), micrometeorites (down to cm), and galactic cosmic rays (down to m) to create the optical surface we see today. We can thus expect that the optical surface of $MU_{69}$ observed with New Horizons is likely only $10^8$-$10^9$ yrs old, having been processed by irradiation and micrometeorite bombardment. Since the colored surfaces of both lobes appeared to be compositionally very uniform to New Horizons (Stern *et al.* 2019, Grundy *et al.* 2020), we can infer that both lobes were compositionally uniform and similar to begin with.

As evidence of potential surface radiolytic effects, it is worth noting that the spectral signature of $CH_3OH$ in the wavelength region in which it was detected with New Horizons (Fig. 1) is subject to subtle modification by radiation processing. Specifically, while the stronger 2.27-μm band is fairly durable against radiation, the weaker band at 2.34 μm is not. Brunetto *et al.* (2005) irradiated films of frozen $CH_3OH$ and water-methanol (1:1) with 30 keV $He^+$ and 200 keV $H^+$ ions. The 2.27-μm band decreased slightly in strength, but the 2.34-μm band eventually disappeared, and new bands at 2.321, 2.355, and 2.37 μm appeared. New molecular bands of CO and $CH_4$ appeared as a result of the irradiation of the methanol-water mix with a dose of 28 eV/16 amu with the 30 keV $He^+$ ions. In the New Horizons LEISA flyby spectrum, the random noise level in the wavelength region of the weaker $CH_3OH$ bands does not allow us to discriminate between unprocessed and processed $CH_3OH$. We note, however, that the original analysis of the spectrum of 5145 Pholus (Cruikshank *et al.* 1998) yielded a good match with $CH_3OH$ (prior to availability of the Brunetto *et al.* 2005 work) and more recent examination of the Pholus spectrum shows, within the constraints of the random noise level, that it is perhaps more consistent with the spectrum of irradiated $CH_3OH$ than it is with pristine $CH_3OH$ in terms of the three new bands that Brunetto *et al.* found.

4.3    One Species Not Well Modeled - $H_2CO$ (Formaldehyde) Ice. As a first product of CO hydrogenation (via H-atom additions) on the way to forming methanol, formaldehyde is a potentially important ice in the makeup of $MU_{69}$. It is also a major radiation degradation product of $CH_3OH$ (Allamandola *et al.* 1988, Schutte *et al.* 1993b). However, the stability of its solid phase ice is difficult to gauge from the extant literature. To quote Fray & Schmitt 2009, "The triple point of $H_2CO$ is located at 155.1 ± 0.3 K, but its pressure is unknown (Table 2). From the data of Spence and Wild (1935), concerning the evaporation equilibrium, we can estimate that the pressure at the triple point is about (4±1) x $10^{-4}$ bar. Lide (2006) reports data only for the evaporation equilibrium. No experimental data concerning the sublimation equilibrium have been found."

Comparative values for methanol are a triple point at 175.5 ± 0.5 K and pressure about (1.8 ± 0.2) x $10^{-6}$ bar, implying that formaldehyde is 200 times more volatile at 20 K lower temperature. Cervinka and Fulem (2017) quote theoretical values for $H_{subl}$ of formaldehyde of 24.4 kJ/mol at 0 K and 26.3 kJ/mol at 155 K, but these are much smaller than the 34.2 ± 1.5 kJ/mol they list as the experimental measure of $H_{subl}$ at 155K, and in all other cases their calculations have either matched experimental values or fallen under them by up to 100%. We thus adopt, with caution, a value for $H_{subl}$ closer to 34 kJ/mol for determining its $P_{sat}$ behavior, and note that this is close to the value of $H_{subl}$ for $NH_3$. Constructing a $P_{sat} = A*e^{(-B/T)}$ curve with $H_{subl} = 34$ kJ/mol and anchored point at 40 Pa and 155 K in Fig. 2, we also find a curve similar to that of $NH_3$. (The triple point (T,P) values of (195.4 K, 0.061 bar) listed by Fray & Schmitt 2009 for $NH_3$ fall nicely on its $P_{sat}$ curve in Fig. 2, giving us confidence in this line of reasoning.) We therefore infer that $H_2CO$ in its pure ice form will have thermal meta-stability over the age of the solar system similar to $NH_3$. This is consistent with its ability to form weak, but finite hydrogen bonds between the carbonyl oxygen and an adjacent aliphatic hydrogen.

One of the difficulties in establishing the sublimation temperature of $H_2CO$ is likely due to the fact that it is highly reactive even at temperatures as low as 40 K (Schutte *et al.* 1993a,b). In pure form it can polymerize into polyoxymethylene (paraformaldehyde) and if other species like $H_2O$, $NH_3$ and $CH_3OH$ are also present, it can polymerize to form a host of compounds having polyoxymethylene backbones with some H atoms substituted for by other functional groups like -$NH_2$, -$CH_3$, and -OH (Schutte *et al.* 1993b). $H_2CO$ is also known to easily cross-link via ion-radical and photon-induced polymerization. Once polymerized, $H_2CO$ is no longer present as a lightweight molecule but is instead associated with structures having much higher molecular weights, i.e., it is no longer a simple ice component, but is instead a complex macromolecule. In this macromolecular form the $H_2CO$ moiety is expected to be very thermally stable vs. sublimation. Polymerized $H_2CO$ could be stored in KBOs and their cometary descendants in this form and would not be expected to be released until exposed to very high temperature excursions (T > 300 K) during inner system passages (e.g. Fray *et al.* 2004).

## 5.    Implications and Predictions.

In the above sections, we presented the results of simple thermodynamic calculations for the stability and lifetime of different ice species on the cold classical KBO body $MU_{69}$. The results for a flattened, 4.6 Gyr old KBO like $MU_{69}$, with obliquity near 90° and spinning around its short axis in a stable $r_h$= 45 AU orbit suggest that only the refractory strongly hydrogen bonded ices ($H_2O$, $CH_3OH$, and HCN, and poly-$H_2CO$) should have remained stable throughout until the present day. The somewhat less refractory and more volatile species $NH_3$ and $CO_2$ could possibly also exist in the cooler "deep interior regions", if further detailed modeling (including realistic ranges of hypervolatiles, short-lived radioactives, and initial temperature conditions) bears out their survival past the first few Myr. All other ice species, especially the hypervolatiles ($N_2$, CO, $CH_4$, Ar, etc.) should be long gone. The importance of refractory hydrogen bonded species like methanol ice in outer solar system bodies is a relatively new concept, as is the rapid removal of hypervolatile ices.

In this section we discuss how these results fit into the bigger picture of other KBOs, the Pluto System, and of the KBOs' dynamical descendants, the Centaurs and Short Period (SP) comets.

5.1 KBOs to Centaurs to Comets. To begin with, we present the state of current observational knowledge of ices on the KBOs, Centaurs, and SP comets. Water and methanol ice, but also ofttimes ammonia and the hypervolatile ices, are known to be present on the largest KBOs (Barucci *et al.* 2008, 2011; Brown 2012). By contrast, the Centaurs show absorptions only due to water ice and methanol (for a few of the most distant objects; in fact the first detection of methanol on a small icy solar system body was for Centaur 5145 Pholus by Cruikshank *et al.* 1998). Correlations of the outgassing activity of 23 Centaurs with their perihelion distance from the Sun led Jewitt (2009) to conclude that the activity of the inner ($r_h < 10$ AU) Centaurs is driven not by CO or $CO_2$ ice sublimation, but instead by crystallization of amorphous water ice and the "squeezing out" of trapped molecules no longer able to fit into the reduced lattice pore space of the newly crystallized ice. SP comet surface spectra do not show any obvious absorption features due to abundant ices, except for the rare small patch of water ice (Sunshine *et al.* 2006; Quirico *et al.* 2015, 2016; Lisse *et al.* 2017). However, their comae, produced largely by water ice sublimation, show an abundant range of icy species (Bockelee-Morvan 2004, Mumma & Charnley 2011) with most species on the order of $0.1 - 1.0$ % of the $H_2O$ gas abundance, with the exception of CO ($0.5 - 25\%$), $CO_2$ (2-12%), and $CH_3OH$ ($0.5 - 5.0\%$) at total relative abundance [CO + $CO_2$ + $CH_3OH$] $\sim 20\%$ (A'Hearn *et al.* 2012).

Given these observational facts, how do we make evolutionary sense of them? **Firstly**, we note that the hypervolatiles CO, $N_2$, and $CH_4$ are commonly present and abundant on the largest KBOs, as well as in the giant planets and their moons – yet our calculations above show (Figs. 4 & 5) that if they were ever incorporated into an $MU_{69}$-sized, near zero-gravity body, they were stably bound at EKB temperatures for less than 1 Myr. This immediately implies that if these largest EKB bodies formed from small KBOs or the same material as smaller KBOs, then (1) they did so within 1 Myr of the PPD mid-plane clearing; or (2) that there was abundant CO, $N_2$, and $CH_4$ gas in the PPD when they became large enough to gravitationally bind these species; or (3) that they are all melted and differentiated through and through, and the surface hypervolatile ices are the result of concentrating the small remnants of the hypervolatiles contained in bulk refractory water phases onto the surface.

All 3 of these Pluto formation scenarios are very plausible given current understanding, and all 3 cases provide important constraints that can be tested and modeled. For instance, in the first two cases, the large hypervolatile-rich bodies have to have formed quickly, via streaming instability/pebble accretion/bottom-up aggregation, within the first few Myr of the Solar System. In the case of Pluto and Charon, this is consistent with the 4 Gyr+ cratering ages found on their surfaces by New Horizons (Robbins *et al.* 2019, Singer *et al.* 2019), and with streaming instability + gravitational instability models of EKB planetesimal formation (McKinnon *et al.* 2019) or pebble accretion (Nesvorny *et al.* 2010, 2019; Lambrechts & Morbidelli 2016) in a pre-migration EKB region of the PPD hundreds of times more massive than today. In the third case, Pluto and Charon would have to be highly differentiated in their interiors; McKinnon *et al.* 2017 argued that differentiated interiors are likely for them from the lack of compressional geological features (and thus the lack of a run of higher and higher density ice phases) seen on the two bodies, while Weaver *et al.* (2016) argued for differentiated proto-Pluto and proto-Charons due to the highly icy nature, and thus highly differentiated nature, of the system's smallest moons, assuming these moons formed during the Pluto-Charon binary-forming impact.

On the other hand, the $N_2/CO \gg 1$ ice abundance ratio found on Pluto's surface (Grundy *et al.* 2016, Protopapa *et al.* 2017, Schmitt *et al.* 2017) is counter to the $N_2/CO \sim 0.10$ abundance ratio expected for these ices trapped in water ice phases; CO with its small but finite dipole moment is much more efficiently trapped in polar water ice than homonuclear, zero dipole $N_2$ from a starting equal mix of the two gases (Yokochi *et al.* 2012; K. Öberg 2019, priv. commun.). However, Kamata *et al.* (2019) argue to the contrary that differential $N_2$ versus CO trapping ability being the case, geochemical and geophysical mechanisms at work on a Pluto surface-devolatilized by a Charon forming impact (and subsequently re-volatilized from deep interior stores of hypervolatiles) that could allow for $N_2$ re-concentration versus CO on its surface.

5.2    No "Top-Down" KBO formation.  If we turn the problem around, and note that we do not see nebular abundances of $N_2$ and CO in the comets or small KBOs, this argues against these objects forming as large bodies and incorporating everything available in the PPD via gravity stabilization, then collisionally fragmenting and retaining these ices. The limits on $MU_{69}$'s density of 0.2 - 0.8g/cm$^3$ (McKinnon *et al.* 2019), and the density values of ~0.5 g/cm$^3$ for comet nuclei and the smaller KBOs vs the range of 1.5 – 2.3 g/cm$^3$ for the largest KBOs (Lacerda & Jewitt 2007, Brown 2013) also argue against a "top-down" KBO formation scenario.

5.3   Implications for Methanol in the Pluto System. The moons of Pluto are on the same size scale as KBO $MU_{69}$, with somewhat lower expected ambient surface temperatures (~35K) due to their very high albedos ($p_v$ = 0.4 to 0.6 vs $MU_{69}$'s $p_v$ = 0.16, Weaver *et al.* 2016) despite their being closer to the sun by a factor of ~45/35. The resulting $v_{escape}$ and $v_{thermal}$ values still clearly put them in the airless $P_{ambient} \approx 0$ regime (Section 2), and we can expect them to have lost all their hypervolatiles over the age of the solar system, assuming their recondensation and incorporation after the Charon-forming impact event that created them (Canup 2005, McKinnon *et al.* 2017) was possible. On the other hand, methanol ice should be quite stable, as seen on $MU_{69}$, and it is possible the moons contain substantial amounts of it.

Similarly, we can expect Pluto and Charon to have large amounts of stable methanol ice content; methanol ice should be almost as strong and as stable as water ice on these bodies. If the primordial KBO methanol: water ratio is on the order of unity, as implied from the NH flyby of $MU_{69}$, then Pluto & Charon, with rock:ice ratios on the order of unity, should contain copious amounts of methanol "rock" in their interiors (and dissolved methanol in their subsurface oceans, if any exist) that need to be considered when modeling their geological structure and history. A search for exposed methanol ice in the NH LEISA data to verify this assertion is difficult on Pluto, however, due to the abundant methane ice coverage on its surface, as methane shares similar 1-2.5 µsm absorption features to methanol.

5.4   Non-Zero Amounts of Hypervolatiles Observed in Comets. The fact that there are low, but finite hypervolatile species seen in SP comets derived from the EKB, implies that there must be some reservoir for them in these recent EKB escapees. The current crop of SP comets have been in the inner system for < 10$^5$ yr (Levison & Duncan 1994, Lisse 2002), so their deep interiors are still warming from the recent increase in surface insolation (Benkhoff & Huebner 1996, Huebner 2002,

Huebner *et al.* 2006) and the fate of any core ices will be dominated by the interior composition established 4.6 Gyr ago. However, our calculations above show that except for $H_2O$, $CH_3OH$, HCN, or polymerized $H_2CO$ compounds, this reservoir cannot contain hypervolatile ice phases, unless they have deeply buried regions held at ~15 K since the beginning of the Solar System.

This is highly unlikely given that the models of Choi *et al.* 2001, Merk & Prialnik 2006, and Prialnik 2009 show that KBO cores are always as warm or warmer than their surface; the lack of substantial hypervolatile emission from end member objects 45P/Honda-Mrkos-Pajdušáková, 46P/Wirtanen, and 103P/Hartley 2 (small comets near the end of their lives emitting chunks of their cores, A'Hearn *et al.* 2011, DiSanti *et al.* 2017); the lack of hypervolatile emission seen from the recently split comets 73P/SW3 (Dello Russo *et al.* 2007) and 17P/Holmes (Dello Russo *et al.* 2008); and the singularly unique behavior of the one truly hypervolatile dominated comet, Oort Cloud object C/2016 R2 (PanSTARRS) (Biver *et al.* 2018, McKay *et al.* 2019; see below). Some authors, like Mousis *et al.* 2019, even argue that there were never any majority hypervolatile ices in small solar system bodies – they condensed directly out of the PPD as water ice dominated mixtures instead, and that impure crystalline water ice dominated, inner system icy small bodies were able to build the giant planets seen today.

We are thus left to follow Iro *et al.*'s 2003 & Jewitt's 2009 concept (now re-checked by Li *et al.* 2020's new outer Centaur activity survey) that the hypervolatiles and moderately-volatile species in the Centaurs and SP comets are protectively stored in $H_2O$ ice matrices – first at high concentrations in cold (T < 80K) amorphous water ice composites, then in lower concentrations in warmer T > 100K crystalline water ice matrices limited by the maximum interlattice "pore space" trapping capability of the crystallite. This is consistent with the early work of Prialnik *et al.* (1987) who argued that the presence of amorphous ice within the subsurface of comets – inferred from observations of outgassing at surprisingly large heliocentric distances (5.8 - 11.4 AU) and attributed to the annealing of amorphous ice as comets first enter the inner Solar System (Prialnik & Bar-Nun, 1990; 1992; Meech *et al.* 2009) – provides a clear constraint on the maximum parent body temperatures (T < 135 K) experienced over comet lifetimes. It is also consistent with the newer work on mixed water ice phases of Guilbert-Lepoutre & Jewitt (2011) and Guilbert-Lepoutre *et al.* (2016) motivated by ROSETTA studies of comet 67P/Churyumov-Gerasimenko.

$H_2O$ ice reservoirs for remnant icy molecules also explain the seeming disconnect between the strong $CH_3OH$ vs $H_2O$ signature on the surface of $MU_{69}$, the weak $CH_3OH:H_2O$ signature of several Centaurs, and the low $CH_3OH:H_2O$ abundance ratio seen in the atmospheres of the active SP comets. Any $CH_3OH$ (and maybe HCN) ice phases that are stable at $MU_{69}$'s T=42 K temperature in the heart of the EKB sublimes within the few Myrs of time it takes to scatter the KBO past Neptune ($r_h \sim 30$ AU, T $\sim 51$ K), then Uranus ($r_h \sim 19$ AU, T $\sim 64$ K), and into Saturn's dynamical regime ($r_h \sim 9.6$ AU, T $\sim 91$ K; Fig. 4). What is left after this process are the molecules mixed in with $H_2O$ ice, and subject to any changes in the state of the $H_2O$ ice, like the amorphous -> crystalline water ice transitions at T = 80 – 120 K (Blake *et al.* 1991), and water ice sublimation at T > 140 K (Sandford & Allamandola 1993b). (Note that both of these processes proceed at a temperature-dependent rate; Schmitt *et al.* (1989) gives the timescale for the amorphous -> crystalline water ice transition as $\tau = (3 \times 10^{-21}$ yrs) $* \exp^{(E_a/kT)}$ with $E_a/k = 5370$ K, and we show the rate of water ice sublimation/unit area vs temperature in Fig 3. Also note that the inter-lattice volume of crystalline $H_2O$ ice can only hold up to $\sim 17\%$ $CO_2$ or $\sim 20\%$ CO by number, similar to the maximum abundances seen in comets for these species [Bockelee-Morvan *et al.* 2004].)

This line of reasoning would be bolstered by a new, improved telescopic survey of the Centaur population to study the pattern of $CH_3OH$ ice presence vs. heliocentric distance; if correct, then abundant $CH_3OH$ ice should exist only in the outer, more distant and inactive Centaurs. In summary, KBOs like MU69 should be much richer in amorphous water ice and its associated minority ice impurities (like the hypervolatiles, $NH_3$, and $CO_2$), as well as the other stable refractory hydrogen bonded ices ($CH_3OH$, HCN), than SP inner system comets dominated by crystalline water ice phases + impurities.

The presence of water ice reservoirs for remnant volatile ice molecules also makes a strong prediction about $Q_{gas}$, the gas production trends for comets. Species that are sourced from amorphous water ice composites do not need to track $Q_{water}$, only the rate at which amorphous water ice crystallizes to produce solid crystalline water + gaseous minor species. Amorphous water ice minor species can also be present in much greater numbers vs water than the limited amount ($\sim 20\%$) possibly carried in the pore spaces of cubic crystalline water, i.e., as $H_2O$ ices warm through $\sim 80$ K, the rearrangement of $H_2O$ molecules during the conversion of $H_2O$ from one amorphous phase to another will allow some hypervolatiles to escape (Sandford & Allamandola 1988, 1990; Blake *et al.* 1991). Additional hypervolatile loss can occur when the ice is warmed to temperatures that convert the amorphous $H_2O$ ice to its cubic crystalline form (Schmitt *et al.* 1989, Jenneskins & Blake 1996)[44]. It is only upon subsequent sublimation of the cubic crystalline $H_2O$ (as seen for comets inside 3 AU) that the emission rate of remaining (~20% total vs water) minority volatiles will track $Q_{water}$.

5.5    C/2016 R2 (PANSTARRS).   Our physical model for $MU_{69}$'s and Pluto's ices has something to add to the discussion of the recently discovered, highly unusual comet C/2016 R2 (PANSTARRS) (hereafter R2). This is a dynamically old Oort Cloud comet ($P_{orbit} \sim 20,000$ yrs) exhibiting a highly unusual coma gas composition. It outgasses CO, $N_2$, $CH_4$, and $CH_3OH$ at extremely high rates compared to its minuscule to nonexistent water gas emission rate (Biver *et al.* 2018, McKay *et al.* 2019). Its effective radius of ~15 km (or less; McKay *et al.* 2019) is comparable to that of $MU_{69}$, but much smaller than any of the KBOs able to gravitationally retain hypervolatile ices. By all coma gas abundance standards, this comet is acting like a piece of thermally unprocessed KBO that has only recently been warmed up past the methanol ice sublimation temperature. Given the $Q_{gas}$ curves of Fig. 5, we could then suppose it was formed at T< 20K in an optically thick mid-plane region, then ejected into the Oort Cloud to stay at these temperatures until very recently, when it was perturbed onto the orbit we currently see today, where it is undergoing intense sublimation of its hypervolatiles, which as we saw in Section 4.1 is enough to sublimatively cool the body to very low temperatures where water ice is rock-like and inactive sublimation-wise.

---

[4]However, if there is sufficient $CH_3OH$ present to force enclathration at ~120 K, ice loss behavior is more complicated. Conversion to a clathrate will allow the ice to accommodate ~7% $CH_3OH$ relative to $H_2O$ and a combined total of ~14% of other volatile species within the clathrate structure; excess amounts of $CH_3OH$ and other volatiles present will be squeezed out of the clathrate. The resulting phase transition forms a porous clathrate structure from which the excess $CH_3OH$ and other molecules can rapidly sublime (Blake *et al.* 1991). This would result in a 'burst' of emitted $CH_3OH$ and other more volatile species, with the released amounts of each being determined by their excess abundances over that which could be accommodated by the clathrate. Once again, the loss of these volatiles would not track $Q_{water}$, although if excess $CH_3OH$ is present during enclathration, they might track $Q_{methanol}$ for the duration of the 'burst'. After enclathration all the remaining volatiles will be trapped in, and controlled by, the clathrate structure and will be unable to leave until the $H_2O$ clathrate structure sublimes. At this point, the loss of other volatiles would be expected to closely track $Q_{water}$ and be constrained to abundance ratios capped by the clathrate structure.

Is this a viable picture? It takes a major dynamical event to put an object onto an Oort Cloud orbit. KBO-KBO collisions, like the Haumea family formation event or the Pluto-Charon formation event, cannot do this as Pluto's escape velocity ($v_{esc}$) and EKB relative velocities are too low (Canup 2005, Stern *et al.* 2006, Sekine *et al.* 2017). Dynamical scattering by a giant planet with a large $v_{esc}$, as in the initial phases of planetary core formation via planetesimal-aggregation, or during proto-Neptune's later migration through the inner proto-EKB, can do this.[5] The latter case would require Neptune's migration through the inner proto-EKB to have occurred no more than a few Myr after the midplane clears for R2's hypervolatiles to have remained stable (Section 4), contrary to the early Nice models (*e.g.* Gomes *et al.* 2005), but consistent with later Nice scenarios including ejected outer planets (*e.g.* Nesvorny & Morbidelli 2012).

For the sake of completeness, another possibility, that R2 somehow retained significant deep interior hypervolatiles in the inner solar system until being scattered into the Oort Cloud, should also be entertained. Following the work of Choi *et al.* 2001, Merk & Prialnik 2006, and Prialnik 2009, this would require an unusually large overabundance of initial hypervolatiles and/or underabundance of short-lived radionuclides. While this seems very unlikely, given the evidence on hypervolatile-devoid comet interior composition we have from observations of end member objects 45P/Honda-Mrkos-Pajdušáková, 46P/Wirtanen, and 103P/Hartley 2 (small comets near the end of their lives emitting chunks of their cores, A'Hearn *et al.* 2011, DiSanti *et al.* 2017); the lack of hypervolatile emission seen from the recently split comets 73P/SW3 (Dello Russo *et al.* 2007) and 17P/(Dello Russo *et al.* 2008), and the hypervolatile-devoid KBO compositions discussed in this paper, it is perhaps possible that this likelihood is no less extreme than requiring an object to be scattered into the Oort Cloud during the era of giant planet formation/migration. Further modeling work will be required to assess the relative likelihoods of these two scenarios.

---

[5] The main issue with this picture is how many times R2 can pass through perihelion and still remain stable against evaporative disintegration. A 15 km radius object of ~0.5g/cm$^3$ density masses on the order of 7 x 10$^{15}$ kg. At the $Q_{gas} \sim 10^{29}$ molecules/sec outgassing levels seen during R2's current apparition for its current perihelion distance of 2.6 AU it will lose ~ 3 x 10$^7$sec [1 yr]*(28 amu for CO/N$_2$*1.67 x 10$^{-27}$ kg/amu)*1 x 10$^{29}$ molecules/sec ~ 1 x 10$^{11}$ kg. Thus it should be able to endure ~7 x 10$^4$ more of these kind of passages before dissipating, which should take ~2 x 10$^4$ yrs/passage * 7 x 10$^4$ passages = 1.4 x 10$^9$ yrs, or 1.4 Gyrs. It is thus unlikely to have been on its present orbit for more than 1 Gyr. Note that a 10$^{29}$ molecules/sec level of hypervolatile outgassing is reasonable, and can be supported by a 15 km radius object that is losing molar amounts of ice surface every second : (2*pi*r$^2$ cm$^2$)*(2x10$^{15}$/cm$^2$/sec) = 3x10$^{28}$ mol/sec. From section 4.1 and Tables 1 & 2, the evaporation of 1 mole of hypervolatile ice requires ~7.3 kJ of heat, thus the amount of hypervolatile cooling from the emission rate of 10$^{29}$ molecules/sec = 1.6 x 10$^5$ moles/sec is ~ 1.4 x 10$^9$ J/sec, the same order of magnitude, (1-0.9) * $\pi$R$_{nucleus}^2$ * (0.1W/cm$^2$ * (1.0 AU/2.6 AU)$^2$) = 1 x 10$^{10}$ W, as the insolation heating of the Sun is delivering to R2's nucleus (with assumed albedo = 0.90) at 2.6 AU. So the observed mass loss rate of R2 being attributed to hypervolatile sublimation makes rough sense if R2 is feverishly sublimating from its entire sunlit surface during the small portion of each 20,000 yr long orbital cycle where it is intensely heated.

In summary, we find that in the ice-sense, $MU_{69}$ is "more primordial" than SP comets, and as or more primitive as the most distant and inactive Centaurs, like 54598 (18.3 AU), 5145 (Pholus, 18.5 AU), and 52975 (21.1 AU). By contrast, $MU_{69}$, like other small KBOs, has not retained much of the PPD hypervolatile ice species found on the largest KBOs, the unique Oort Cloud comet C/2016 R2 (PANSTARRS), and in the giant planets and their moons, as it was unable to gravitationally bind them versus their thermal instability in the present Edgeworth-Kuiper Belt. What it does contain of these species is likely bound up as minor impurities in more thermally stable water ice phases.

## 6.  Conclusions.

In this paper we have found, from simple thermodynamic arguments independent of detailed formation scenario assumptions (e.g., formation location in the proto-solar nebula or the PPD, amount of $Al^{26}$ incorporated & when, interior detailed structure), that New Horizons should not have seen any typical cometary icy material on the surface of $MU_{69}$ other than the highly refractory ices $H_2O$, $CH_3OH$, HCN, and poly-$H_2CO$. These are the maximal possible set for the current low-level insolation conditions. Any positive temperature excursions, due to nearby passing hot stars, nearby supernovae, impacts, meteorite gardening, etc. could only remove more ice, and reduce this possible set. The New Horizons science team has already announced the detection and $CH_3OH$ and evidence for $H_2O$ on this object (Lisse *et al.* 2017, Stern *et al.* 2019, Grundy *et al.* 2020). We predict that further absorption feature detections, if any, will be due to an HCN or poly-$H_2CO$ based ice species. In having evidence for these additional non-water ice phases from the NH flyby spectral mapping of its surface, MU69 appears akin to the most distant inactive Centaurs like 54598, 5145 (Pholus), and 52975. Because of this last point, we suggest a new survey of the distant Centaur and small KBO populations to search for methanol ices.

Compared to an inner system SP comet dominated by crystalline ice phases with limited interlattice carrying volume for minor impurities, our thermal stability analysis suggests that KBOs like MU69 should be much richer in amorphous water ice and its associated minority ice impurities (like the hypervolatiles, $NH_3$, and $CO_2$), as well as the other stable refractory hydrogen bonded ices ($CH_3OH$, HCN). On the other hand, $MU_{69}$ should have lost all of its original hypervolatile (CO, $N_2$, $CH_4$) majority ice phases, just as the SP comets have. Current models of our PPD, coupled with ALMA exo-PPD observations, lead us to the conclusion that $MU_{69}$ "formed in the dark" in the EKB region, unable to see the nascent, variable, highly luminous YSO/TTauri Sun (Briceno *et al.* 2001, Thrane *et al.* 2006), in an optically thick mid-plane. It was thus able to initially incorporate hypervolatile ices at a local T < 20 K, but that these were quickly lost from the body within the first 1-10 Myr due to the combined action of short-lived radionuclide decay and surface warming upon PPD mid-place clearing. By contrast, Pluto was able to retain its original primordial volatiles via gravitational capture into an exobase. Finally, the advent of the uniquely hypervolatile-dominated object comet C/2016 R2 (PanSTARRS) shows us what a truly primordial, hypervolatile ice rich object behaves like upon warming in the inner system. Its uniqueness compared to thousands of other comets is striking, and highlights what has not been seen in recently fragmented, split, or hyperactive near-dead cometary cores, nor in $MU_{69}$ – evidence for deep down core hypervolatile ices. This leads us to conclude that comet R2 was likely placed onto an Oort Cloud orbit very quickly after its formation in the optically thick PPD mid-plane, by scattering from a nascent giant planet or by proto-Neptune as it was migrating.

## 7.    Acknowledgements.

The authors would like to thank NASA for financial support of the New Horizons project that funded this study, and we thank the entire New Horizons mission team for making the success of the flyby and its fantastic data return possible.

## 8.    References

Abell, P.A. *et al.* 2005, *Icarus* **179**, 174

A'Hearn, M.F. *et al.* 2011, *Science* **332**, 1396

A'Hearn, M.F.,  Feaga, L.M., Keller, H.U. *et al.* 2012, *Astrophys. J* **758**, 29

Allamandola, L. J., Sandford, S. A., & Valero, G. 1988,  *Icarus* **76**, 225

Allamandola, L. J., Sandford, S. A., Tielens, A. G. G. M., & Herbst, T. M. 1992,  *Astrophys. J.* **399**, 134

Anderl, S., Maret, S., Cabrit, S. et al. 2016, Astron. Astrophys. 591, A3  CO/CH3OH snow lines

Bar-Nun, A., Herman, G., Rappaport, M.L. & Laufer, D. 1985, *Icarus* **63**, 317-332

Bar-Nun, A., Dror, J. , Kochavi, E., Laufer, D.  1987, *Phys. Rev. B* **35**, 2427

Barucci M.A., Brown M.E., Emery J.P., & Merlin F. 2008, "Composition and Surface Properties of Transneptunian Objects and Centaurs", in *The Solar System Beyond Neptune*, eds. M.A. Barucci, H. Boehnhardt, D.P. Cruikshank, & A. Morbidelli. 143–160

Barucci M.A., Alvarez-Candal A., Merlin F., Belskaya I.N., de Bergh C., *et al.* 2011, *Icarus* **214**, 297–307

Benkhoff, J. & Huebner, W.F. 1996, *Planet. & Space Sci* **44**, 1005-1013

Betrand, T. *et al.* 2019, *Icarus* **329**, 148-165

Bieler, A. *et al.* 2015, "Abundant molecular oxygen in the coma of comet 67P/Churyumov-Gerasimenko", *Nature* **526**, 678–681

Biver, N. *et al.* 2018, *Astron Astrophys.* **619**, A127

Blake, D., Allamandola, L., Sandford, S., Hudgins, D., & Freund, F. (1991). Clathrate Hydrate Formation in Amorphous Cometary Ice Analogs in Vacuo. *Science* **254**, 548-551

Bockelée-Morvan, D., Crovisier, J., Mumma, M. J., & Weaver, H. A. 2004, "The Composition of Cometary Volatiles", in *Comets II*, M. C. Festou, H. U. Keller, and H. A. Weaver (eds.), University of Arizona Press, Tucson, 745 pp., p. 391-423

Briceno, C., *et al.* 2001, *Science* **291**, 93

Brown, M. E., Burgasser, A. J. & Fraser, W.C. 2011, "The surface composition of large Kuiper Belt Object 2007 OR10", *Astrophys. J* **738**, L26

Brown, M.E. 2012, *Ann Rev Earth & Planet Sci* **40**, 467-494

Brown, M.E. 2013, *Astrophys. J Lett*, **778**, 2

Brunetto, R., G. A. Baratta, M. Domingo, & G. Strazzulla 2005, "Reflectance and transmittance spectra (2.2-2.4 μm) of ion-irradiated frozen methanol", *Icarus* **174**, 226-232.

Canup, R. M. 2005, *Science* **307**, 546–550

Castillo-Rogez J. C., Matson D. L., Sotin C., *et al.* 2007, *Icarus* **190**, 179

Catling, D. C., & Zahnle, K. J. 2009, *SciAm* **300**, 36

Catling, D. C., & Zahnle, K. J. 2017, *Astrophys. J* **843**, 122

Cervinka, and Fulem 2017, *J. Chem. Theory Comput.* **13**, 2840−2850

Choi, Y.-J., Cohen, M., Merk, R., Prialnik, D. 2002, *Icarus* **160**, 300-312

Cruikshank, D. P., Roush, T. L., Bartholomew, M. J., *et al.* 1998, *Icarus* **135**, 389-407

Dello Russo, N., Vervack, R.J., Jr., Weaver, H.A., *et al.* 2007, *Nature* **448**, 172

Dello Russo, N., Vervack, R.J., Jr., Weaver, H.A., *et al.* 2008, *Astrophys. J.* **680**, 793

DiSanti, M.A., Bonev, B.P., Dello-Russo, N., *et al.* 2017, *Astron. J* **154**, 246

Earle, A. *et al.* 2017, *Icarus* **287**, 37-46.

Fray, N., Be´nilan, Y., Cottin, H., & Gazeau, M.-C. 2004, *JGR* **109**, E07S12

Fray, N. & Schmitt, B. 2009, *Planet. Space Sci* **57**, 2053–2080

Gladstone, G.R. *et al.* 2016, *Science* **351**, 8866

Glein, C.R. & Waite, J.H. 2018, *Icarus* **313**, 79-92

Gounelle, M., Morbidelli, A., Bland, P. A., *et al.* 2008, "Meteorites from the Outer Solar System?", in *The Solar System Beyond Neptune*, M. A. Barucci, H. Boehnhardt, D. P. Cruikshank, and A. Morbidelli (eds.), University of Arizona Press, Tucson, 592 pp., p.525-541

Grundy, W.M. *et al.* 2016, *Science* **351**, 9189

Grundy, W.M., M. K. Bird, M.K., Britt, D.T. *et al.* 2020, *Science* **367**, eaay3705

Guilbert-Lepoutre, A., & Jewitt, D. 2011, *Astrophys. J* **743**, 31

Guilbert-Lepoutre, A., Rosenberg, E. D., Prialnik, D., & Besse, S. 2016, *MNRAS* **462**, S146 – S155

Huebner, W.F. 2002, *Earth, Moon, & Planets* **89**, 179-195

Huebner, W.F., Benkhoff, J., Capria, M.-T., *et al.* 2006, "Heat and Gas Diffusion in Comet Nuclei", SR-004, ISBN 1608-280X. Published for the International Space Science Institute, Bern, Switzerland, by ESA Publications Division, Noordwijk, The Netherlands, June 2006.

Iro, N. , Gautier, D., Hersant, F., Bockelée-Morvan, D., Lunine, J. 2003, *Icarus* **161**, 511-532

Jenniskens, P. & Blake, D.F. 1996, Astrophys J 473, 1104

Jewitt, D.C. 2009, *Astron J* **137**, 4296–4312

Keeney, B. A., Stern, S. A., A'Hearn, M. F., *et al.* 2017, *MNRAS* **469**, S158

Krijt, S., Schwarz, K.R., Bergin, E.A., & Ciesla, F.J. 2018, Astrophys J 864, 78

Krijt, S., Bosman, A.D., Zhang, K. et al. 2020, "CO Depletion in Protoplanetary Disks: A Unified Picture Combining Physical Sequestration and Chemical Processing", arXiv:2007.09517

Krot, A. N.., Makide, K., Nagashima, K., *et al.* 2012, *Meteoritics & Planetary Sci* **47**, 1948-1979.

Lacerda, P. & Jewitt, D.C. 2007, *Astron. J* **133**, 1393

Lacy, J. H., Faraji, H., Sandford, S. A., & Allamandola, L. J. (1998). Unraveling the 10 μm 'Silicate' Feature of Protostars: The Detection of Frozen Interstellar Ammonia. *Astrophys. J. (Letters)* **501**, L105-L109.

Lambrechts, M. & Morbidelli, A. 2016, "Reconstructing the Size Distribution of the Small Body Population in the Solar System", AAS/Division for Planetary Sciences Meeting Abstracts No. **48**, ID.105.08

Langmuir, I. 1916, *Phys. Rev.* **8**, 149

Laufer, D., Bar-Nun, A. & Greenberg, A.N. 2017, *MNRAS* **469**, S818

Lesniak, M.V. & Desch, S.J. 2011, Astrophys J 740, 118

Levison, H.F. & Duncan, M.J. 1994, *Icarus* **108**, 18-36

Li, J., Jewitt, D., Mutchler, M., Agarwal, J. & Weaver, H. 2020, Astron J 159, 209

Lisse, C.M. 2002, *Earth, Moon, & Planets* **90**, 497-506

Lisse, C.M., Bar-Nun, A., Laufer, D., Belton, M.J.S., Harris, W.M., Hsieh, H.H., Jewitt, D.C. 2013, "Cometary Ices", chapter in *"The Science of Solar System Ices"*, Astrophys & Space Sci Library **356**, ISBN 978-1-4614-3075-9. Springer Science+Business Media New York, 455

Lisse, C.M. *et al.* 2017, *Astron. J.* **154**, 5

Loomis, R.A., Öberg, K.I., Andrews, S.M. et al. 2020, Astrophys J 893, 101

Mall, U., Altwegg, K., Balsiger, H. *et al.* 2016, "High-Time Resolution in situ Investigation of Major Cometary Volatiles around 67P/C-G at 3.1–2.3 AU Measured with ROSINA-RTOF", *Astrophys. J.* **819**, 126

McKay, A.J., DiSanti, M.A., Kelley, M.S.P., *et al.* 2019, *Astron. J.* **158**, 128

McKinnon, W.B., Sterm S.A., Weaver, H.A., *et al.* 2017, *Icarus* **287**, 2-11

McKinnon, W. B., Richardson, D. C., Marohnic, J. C. *et al.* 2020, *Science* **367**, aay6620

Merk, R. & Prialnik, D. 2006, *Icarus* **183**, 283-295 Moore, J.M., *et al.* 2018, *Icarus* **300**, 129-144

Miller, G. A. & Carpenter, D. K. 1964*, J. Chem. Eng. Data* **9**, pp 371–373

Moore, J.M., Howard, A.D., Umurhan, O.M. *et al.* 2018, Icarus 300, 129-144, doi:10.1016/j.icarus.2017.08.031

Moore, J.M. *et al.* 2019, "Scarp Retreat on MU69: Evidence and Implications for Composition and Structure", EPSC Abstracts 13, EPSC-DPS2019-50-1

Mousis, O., Ronnet, T., Lunine, J.I., *et al.* 2018, *Astrophys J. Lett*, **865**, L11

Mousis, O., Ronnet, T., & Lunine, J.I. 2019, *Astrophys J.* **875**, 9

Mumma, M.J. & Charnley, S. 2011, *Ann. Rev Astron Astrophys.* **49**, 471-524

Nesvorný, D., Youdin, A.N., & Richardson, D.C. 2010, *Astron. J.* **140**, 785-793

Nesvorny, D. & Morbidelli, A. 2012, *Astron J* **144**, 4

Nesvorny, D., Li, R., Youdin, A. N., Simon, J. B., & Grundy, W. M. 2019. *Nature Astron.* **3**, 808-812

Nuth, J.A., Johnson, N.M., & Manning, S. 2008, *Astrophys J Lett* **673**, L225

Öberg, K.I., Guzmán, V.V., Merchantz, C.J. *et al.* 2017, *Astrophys J* **839**, 1 (H₂CO Distribution and Formation in the TW HYA Disk)

Prialnik, D., Bar-Nun, A. & Podolak, M. 1987, *Astrophys. J.* **319**, 992

Prialnik, D. & Bar-Nun, A. 1990, *Astrophys. J.* **355**, 281

Prialnik, D. 2002, *EM&P* **89**, 27-52

Prialnik, D., Benkhoff, J., & Podolak, M. 2004, *"Modeling the Structure and Activity of Comet Nuclei"*, chapter in *Comets II*, M. C. Festou, H. U. Keller, and H. A. Weaver (eds.), University of Arizona Press, Tucson, 745 pp., 359-387

Protopapa, S. *et al.* 2017, *Icarus* **287**, 218-228

Quirico, E., Douté, S., Schmitt, B., de Bergh, C., *et al.* 1999, "Composition, Physical State and Distribution of Ices at the Surface of Triton", *Icarus* **139**, 159-178

Quirico, E. *et al.* 2015, *EPSC Abstracts* **10**, EPSC2015-621

Quirico, E. *et al.* 2016, *Icarus* **272**, 32

Reuter, D. C., Stern, S. A., Scherrer, J., *et al.* 2008, "Ralph: A Visible/Infrared Imager for the New Horizons Pluto/Kuiper Belt Mission", *Space Science Reviews* **140**, 129-154.

Robbins, S.J., Beyer, R.A., Spencer, J.R., *et al.* 2019, *JGR: Planets* **124**, 155-174

Sandford, S. A., & Allamandola, L. J. 1988, The condensation and vaporization behavior of H2O:CO ices and implications for interstellar grains and cometary behavior. *Icarus* **76**, 201-224.

Sandford, S. A., & Allamandola, L. J. 1990, The Volume and Surface Binding Energies of Ice Systems Containing CO, CO₂, and H₂O, *Icarus* **87**, 188-192.

Sandford, S.A. & Allamadola, L.J. 1993a, *Icarus* **106**, 478-488

Sandford, S.A. & Allamadola, L.J. 1993b, *Astrophys J.* **417**, 815-825

Sandford, S.A., Allamadola, L.J., Geballe, T.R. 1993, *Science* **262**, 400 – 402

Schaller, E.L. & Brown, M. E. 2007, *Astrophys. J* **659**, L61

Schenk, P.M., Singer, K., Beyer, R. *et al.* 2020, *Icarus* **113834**, doi:10.1016/j.icarus.2020.113834 https://www.sciencedirect.com/science/article/abs/pii/S0019103520302153

Schmitt, B., Espinasse, S., Grim, R. J. A., Greenberg, J. M., & Klinger, J. 1989, in *Physics and Mechanics of Cometary Materials* (ESA SP-302; Noordwijk:ESA Publications Division), 65

Schmitt, B., 1992, "Interrelations Between Physics and Dynamics for Minor Bodies in the Solar System", Comptes Rendus de la 15ieme Ecole de Printemps d'Astrophysique de Goutelas, France, 29 Avril - 4 Mai 1991. Edited by Daniel Benest and Claude Froeschle. Paris: Societe Francaise des Specialistes d'Astronomie (SFSA), 1992., p.265

Schmitt, B. *et al.* 2017, *Icarus* **287**, 229-260

Schutte, W. A., Allamandola, L. J., & Sandford, S. A. (1993a). Formaldehyde and Organic Molecule Production in Astrophysical Ices at Cryogenic Temperatures. *Science* **259**, 1143-1145.

Schutte, W. A., Allamandola, L. J., & Sandford, S. A. (1993b). An Experimental Study of the Organic Molecules Produced in Cometary and Interstellar Ice Analogs by Thermal Formaldehyde Reactions. *Icarus* **104**, 118-137.

Sekine, Y., Genda, H., Kamata, S. & Funatsu, T. 2017, The Charon-forming giant impact as a source of Pluto's dark equatorial regions. *Nature Astronomy* **1**, 0031

Shock, E. L. & McKinnon, W. B. 1993, *Icarus* **106**, 464-477

Singer, K., McKinnon, W. B., Gladman, B., *et al.* 2019, *Science* **363**, 955-959

Spencer, J. R., Stern, S. A., Moore, J. M. *et al.* 2020, *Science* **367**, aay3999

Steckloff, J.K., Johnson, B.C., Bowling, T. *et al.* 2015, *Icarus* **258**, 430 – 437

Steckloff, J.K., Lisse, C.M., Safrit, T.K., Bosh, A.S., Lyra, W. 2020, *ibid (Icarus, this issue)*

Stern, S.A. 2003, *Nature* **424**, 639-642

Stern, S. A. *et al. 2006, Nature* **439**, 946

Stern, S.A. *et al.* 2015, *Science* **350**, 1815

Stern, S.A., Weaver, H.A., Spencer, J.R. *et al.* 2019, *Science* **364**, 9771

Sunshine, J.S. *et al.* 2006, *Science* **311**, 1453-1455

Tang, H. & Dauphas, N. 2015, *Astrophys. J.* **802**, 22

Thrane, K., Bizzarro, M. & Baker, J.A. 2006, *Astrophys. J* **646**, L159–L162

Weaver, H. A., Buie, M.W., Buratti, B.J. *et al.* 2016, *Science* **351**, 30

Yamamoto, T., Nakagawa, N., & Fukui, Y. 1983, *Astron Astrophys.* **122**, 171-176

Yokochi, R., Marboeuf, U., Quirico, E., & Schmitt, B. 2012, *Icarus* **218**, 760-770

Young, L.A. *et al.* 2018, *Icarus* **300**, 174-199

# Supplementary Online Material for ICARUS_2019_521, "On the Origin & Thermal Stability of KBO 2014 MU69's and Pluto's Ices"

Appendix : Experimental Laboratory Saturation Vapor Pressure Fits

These two tables present the best-fit parameters compiled/used by Prialnik et al. (2004) and Fray & Schmitt (2009) to describe the vapor pressure $P_{sat}$ versus temperature T behavior of ices found in solar system bodies. We reproduce them here both to record the values used in making the vapor pressure versus temperature curves used in our manuscript's analysis and as a service to the reader in making their own future analyses.

The 2-parameter fits of Prialnik et al. (2004; Table 1) assume ice Heats of Vaporization ($\Delta H_{vap}$'s) that are independent of temperature and use data collected at/near room temperature, and so are most accurate at/near room temperature and less so at low cryogenic temperatures. These parameters are most familiar to the cometary community of researchers at the time of this writing.

Table 1 – Prialnik et al. 2004 Empirical $P_{sat} = A*exp^{(-B/T)}$ Parameter Values Assuming $\Delta H_{vap}$ = Constant

| Ice Component | Formula | A Value $(10^{10}$ Nm$^{-2})$ | B Value (Kelvin) |
|---|---|---|---|
| Water | $H_2O$ | 356. | 6141.667 |
| Carbon Monoxide | CO | 0.12631 | 764.16 |
| Carbon Dioxide | $CO_2$ | 107.9 | 3148 |
| Methane | $CH_4$ | 0.597 | 1190.2 |
| Propyne | $C_3H_4$ | 3.417 | 3000 |
| Propadine | $C_3H_4$ | 2.382 | 2758 |
| Ethane | $C_2H6$ | 0.459 | 1938 |
| Methanol | $CH_3OH$ | 8.883 | 4632 |
| Hydrogen Cyanide | HCN | 3.8665 | 4024.66 |
| Hydrogen Sulphide | $H_2S$ | 1.2631 | 2648.42 |
| Ammonia | $NH_3$ | 61.412 | 3603.6 |
| Acetylene | $C_2H_2$ | 9.831 | 2613.6 |

The more extensive compendium of laboratory measurements compounded by Fray & Schmitt (2009; Table 2), produced by an extensive search & comparison of the existing literature for laboratory data, expands the list of ices studied to encompass most of the materials expected in outer solar system icy bodies. It also includes the results of heat of vaporization and vapor pressure measurements at temperatures ranging from 10 to 300 K for the candidate ices, without assuming that $\Delta H_{vap}$ is constant with Temperature. The resulting polynomial fits of $ln(P_{sat})$ in $1/T$ (i..e, $ln(P_{sat})$ = $A_0 + \sum A_i/T^i$ )were tabulated and compared to those of other works (like Huebner 2006) in their work with good results. These parameters, along with the subset used by Schaller & Brown (2007), are most familiar to the KBO/TNO community of researchers.

We recommend the use of the Fray and Schmitt 2009 values as being more completely researched and more accurate throughout the temperature range of interest for icy solar system bodies, especially at low temperatures below 50K, and this is what we have done in our analysis. On the other hand the Prialnik et al. 2004 formulation is much easier to use for a quick order of magnitude estimate of the $P_{sat}$ vs Temperature behavior: the A coefficient tells one quickly the relative volatility of a species, while the B coefficient divided by 35 gives one a quick approximate temperature where $P_{sat}$ climbs steeply from negligible to rapidly sublimating for an ice.

Table 2 - Coefficients of the polynomials of extrapolations $\ln(P_{sat}) = A_0 + \sum A_i/T^i$ , from Fray & Schmitt 2009

| Species | Polynomial Designation | A0 | A1(K) | A2(K²) | A3(K³) | A4(K⁴) | A5(K⁵) | A6(K⁶) |
|---|---|---|---|---|---|---|---|---|
| $H_2O$ | $H_2O$-1 | - See below for special treatment of water ice vapor pressure - | | | | | | |
| $O_2$ | $O_2$-1 | $1.541\times10^{+1}$ | $-1.148\times10^{+3}$ | $3.349\times10^{+2}$ | $6.021\times10^{+1}$ | 0 | 0 | 0 |
| | $O_2$-2 | $1.335\times10^{+1}$ | $-1.012\times10^{+3}$ | $-2.971\times10^{+3}$ | $2.926\times10^{+4}$ | 0 | 0 | 0 |
| | $O_2$-3 | $1.018\times10^{+1}$ | $-8.035\times10^{+2}$ | $-7.080\times10^{+3}$ | $7.553\times10^{+4}$ | 0 | 0 | 0 |
| $O_3$ | $O_3$-1 | $1.746\times10^{+1}$ | $-2.352\times10^{+3}$ | | | | | |
| CO | CO-1 | $1.043\times10^{+1}$ | $-7.213\times10^{+2}$ | $-1.074\times10^{+4}$ | $2.341\times10^{+5}$ | $-2.392\times10^{+6}$ | $9.478\times10^{+6}$ | 0 |
| | CO-2 | $1.025\times10^{+1}$ | $-7.482\times10^{+2}$ | $-5.843\times10^{+3}$ | $3.939\times10^{+4}$ | 0 | 0 | 0 |
| $CO_2$ | $CO_2$-1 | $1.476\times10^{+1}$ | $-2.571\times10^{+3}$ | $-7.781\times10^{+4}$ | $4.325\times10^{+6}$ | $-1.207\times10^{+8}$ | $1.350\times10^{+9}$ | 0 |
| | $CO_2$-2 | $1.861\times10^{+1}$ | $-4.154\times10^{+3}$ | $1.041\times10^{+5}$ | 0 | 0 | 0 | 0 |
| $CH_3OH$ | $CH_3OH$-1 | $1.918\times10^{+1}$ | $-5.648\times10^{+3}$ | 0 | 0 | 0 | 0 | 0 |
| | $CH_3OH$-2 | $1.706\times10^{+1}$ | $-5.314\times10^{+3}$ | 0 | 0 | 0 | 0 | 0 |
| HCOOH | HCOOH-1 | $2.189\times10^{+1}$ | $-7.213\times10^{+3}$ | | | | | |
| | HCOOH-2 | $2.164\times10^{+1}$ | $-6.942\times10^{+3}$ | $-6.579\times10^{+4}$ | $3.316\times10^{+6}$ | $-6.004\times10^{+7}$ | 0 | 0 |
| $CH_4$ | $CH_4$-1 | $1.051\times10^{+1}$ | $-1.110\times10^{+3}$ | $-4.341\times10^{+3}$ | $1.035\times10^{+5}$ | $-7.910\times10^{+5}$ | 0 | 0 |
| $C_2H_2$ | $C_2H_2$-1 | $1.340\times10^{+1}$ | $-2.536\times10^{+3}$ | | | | | |
| $C_2H_4$ | $C_2H_4$-1 | $1.540\times10^{+1}$ | $-2.206\times10^{+3}$ | $-1.216\times10^{+4}$ | $2.843\times10^{+5}$ | $-2.203\times10^{+6}$ | 0 | 0 |
| $C_2H_6$ | $C_2H_6$-1 | $1.511\times10^{+1}$ | $-2.207\times10^{+3}$ | $-2.411\times10^{+4}$ | $7.744\times10^{+5}$ | $-1.161\times10^{+7}$ | $6.763\times10^{+7}$ | 0 |
| $C_6H_6$ | $C_6H_6$-1 | $1.735\times10^{+1}$ | $-5.663\times10^{+3}$ | | | | | |
| HCN | HCN-1 | $1.393\times10^{+1}$ | $-3.624\times10^{+3}$ | $-1.325\times10^{+5}$ | $6.314\times10^{+6}$ | $-1.128\times10^{+8}$ | 0 | 0 |
| $HC_3N$ | $HC_3N$-1 | $1.301\times10^{+1}$ | $-4.426\times10^{+3}$ | | | | | |
| $C_2N_2$ | $C_2N_2$-1 | $1.653\times10^{+1}$ | $-4.109\times10^{+3}$ | | | | | |
| $C_4N_2$ | $C_4N_2$-1 | $1.909\times10^{+1}$ | $-6.036\times10^{+3}$ | | | | | |
| $N_2$ | $N_2$-1 | $1.240\times10^{+1}$ | $-8.074\times10^{+2}$ | $-3.926\times10^{+3}$ | $6.297\times10^{+4}$ | $-4.633\times10^{+5}$ | $1.325\times10^{+6}$ | 0 |
| | $N_2$-2 | $8.514$ | $-4.584\times10^{+2}$ | $-1.987\times10^{+4}$ | $4.800\times10^{+5}$ | $-4.524\times10^{+6}$ | 0 | 0 |
| $NH_3$ | $NH_3$-1 | $1.596\times10^{+1}$ | $-3.537\times10^{+3}$ | $-3.310\times10^{+4}$ | $1.742\times10^{+6}$ | $-2.995\times10^{+7}$ | | |
| | NO-1 | $1.691\times10^{+1}$ | $-2.016\times10^{+3}$ | 0 | 0 | 0 | 0 | 0 |
| | NO-2 | $1.2352\times10^{+2}$ | $-4.7607\times10^{+4}$ | $7.7292\times10^{+6}$ | $-6.4950\times10^{+8}$ | $2.7061\times10^{+10}$ | $-4.4739\times10^{+11}$ | 0 |
| $N_2O$ | $N_2O$-1 | $1.622\times10^{+1}$ | $-2.971\times10^{+3}$ | | | | | |
| | $N_2O$-2 | $6.5664$ | $-1.2711\times10^{+3}$ | $-6.6835\times10^{+5}$ | $4.4959\times10^{+7}$ | $-1.0967\times10^{+9}$ | 0 | 0 |
| $H_2S$ | $H_2S$-1 | $1.298\times10^{+1}$ | $-2.707\times10^{+3}$ | 0 | 0 | 0 | 0 | 0 |
| | $H_2S$-2 | $8.933$ | $-7.260\times10^{+2}$ | $-3.504\times10^{+5}$ | $2.724\times10^{+7}$ | $-8.582\times10^{+8}$ | 0 | 0 |
| $SO_2$ | $SO_2$-1 | $1.560\times10^{+1}$ | $-3.5.08\times10^{+3}$ | $-9.401\times10^{+4}$ | $4.152\times10^{+6}$ | $-6.946\times10^{+7}$ | | |
| | $AsH_3$-1 | $1.176\times10^{+1}$ | $-2.382\times10^{+3}$ | 0 | 0 | 0 | 0 | 0 |
| Ne | Ne-1 | $9.886$ | $-2.699\times10^{+2}$ | $1.283\times10^{+2}$ | $-1.624\times10^{+2}$ | 0 | 0 | 0 |
| | Ne-2 | $1.061\times10^{+1}$ | $-3.086\times10^{+2}$ | $9.860\times10^{+2}$ | $-9.069\times10^{+3}$ | $3.514\times10^{+4}$ | 0 | 0 |
| Ar | Ar-1 | $1.069\times10^{+1}$ | $-8.932\times10^{+2}$ | $-3.567\times10^{+3}$ | $6.574\times10^{+4}$ | $-4.280\times10^{+5}$ | | |
| Kr | Kr-1 | $1.077\times10^{+1}$ | $-1.223\times10^{+3}$ | $-8.903\times10^{+3}$ | $2.635\times10^{+5}$ | $-4.260\times10^{+6}$ | $3.575\times10^{+7}$ | $-s1.210\times10^{+8}$ |
| Xe | Xe-1 | $1.098\times10^{+1}$ | $-1.737\times10^{+3}$ | $-1.332\times10^{+4}$ | $4.349\times10^{+5}$ | $-7.027\times10^{+6}$ | $4.447\times10^{+7}$ | 0 |

The special case of Water ($H_2O$). As a unique case due to its prevalence, its highly hydrogen bonded nature, and its phase changes at relatively high temperatures (for ices), water ice has been studied most extensively by many researchers over the last century and thus modeled differently in the literature. For water in "Astrophysical

Applications" Fray & Schmitt 2009 recommend the special (i.e., different than the $\ln(P_{sat}) = A_0 + \sum A_i/T^i$ of Table 2) treatment presented by Feistel and Wagner (2007):

$$\ln(P_{subl}(T)/P_{triple}) = 3/2 \ln(T/T_{triple}) + (1 - T_{triple}/T)\, \eta(T/T_{triple})$$

where

$P_{triple}$ = Pressure of Triple Point = $(6.116577 +/- 0.0001) \times 10^{-3}$ bar
$T_{triple}$ = Temperature of Triple Point = 273.16 K

and

$$\eta(T/T_{triple}) = \sum e_i (T/T_{triple})^i$$

with

| i | $e_i$ |
|---|---|
| 0 | 20.9969665107897 |
| 1 | 3.72437478271362 |
| 2 | -13.9205483215524 |
| 3 | 29.6988765013566 |
| 4 | -40.1972392635944 |
| 5 | 29.7880481050215 |
| 6 | -9.13050963547721 |

## Appendix References

Fray, N. & Schmitt, B. 2009, *Planet. Space Sci* **57**, 2053–2080

Feistel, R., Wagner, W., 2007. "Sublimation Pressure and Sublimation Enthalpy of H2O Ice Ih Between 0 and 273.16K", Geochim. Cosmochim. Acta 71, 36

Huebner, W.F., Benkhoff, J., Capria, M.-T., *et al.* 2006, "Heat and Gas Diffusion in Comet Nuclei", SR-004, ISBN 1608-280X. Published for the International Space Science Institute, Bern, Switzerland, by ESA Publications Division, Noordwijk, The Netherlands, June 2006.

Prialnik, D., Benkhoff, J., & Podolak, M. 2004, *"Modeling the Structure and Activity of Comet Nuclei"*, chapter in *Comets II*, M. C. Festou, H. U. Keller, and H. A. Weaver (eds.), University of Arizona Press, Tucson, 745 pp., 359-387

Schaller, E.L. & Brown, M. E. 2007, *Astrophys. J* **659**, L61